\documentclass[10pt,twoside,a4paper,twocolumn]{article}

\usepackage[pdftex]{graphicx}
\usepackage{amsmath}
\usepackage{url}

\date{}

\usepackage{amssymb}
\usepackage{amsmath}
\usepackage{multirow}

\newcommand{\tv}{\ensuremath{\text{TGV}}}

\newcommand{\fiin}{\ensuremath{\mathcal{I}}}

\renewcommand{\ln}{\ensuremath{\mathrm{ln}}}

\renewcommand{\eqref}[1]{Eq. \ref{#1}}

\renewcommand{\(}{\ensuremath{\left(}}
\renewcommand{\)}{\ensuremath{\right)}}
\renewcommand{\[}{\ensuremath{\left[}}
\renewcommand{\]}{\ensuremath{\right]}}
\newcommand{\absl}{\ensuremath{\left|}}
\newcommand{\absr}{\ensuremath{\right|}}
\newcommand{\E}{\ensuremath{\text{E}}}
\newcommand{\cov}{\ensuremath{\text{cov}}}
\newcommand{\dqe}{\ensuremath{\text{DQE}}}
\newcommand{\snrin}{\ensuremath{\text{SNR}_{\text{in}}}}
\newcommand{\snrout}{\ensuremath{\text{SNR}_{\text{out}}}}
\newcommand{\crb}{\ensuremath{\text{AOC}}}

\newcommand{\popt}{\ensuremath{p_{\text{opt}}}}
\newcommand{\tr}{\ensuremath{\text{tr}}}
\newcommand{\av}[1]{\ensuremath{\left\langle #1 \right\rangle}}
\newcommand{\LA}{\ensuremath{\mathcal{L}_{\text{A}}}}

\newcommand{\mse}{\ensuremath{\text{MSE}}}
\newcommand{\elctr}{\ensuremath{\text{e}^{-}}}

\newcommand{\sipidi}{PIX$_{\text{Di}}$}
\newcommand{\sipifr}{PIX$_{\text{Fr}}$}
\newcommand{\adfcs}{ADF$_{\text{CS}}$}
\newcommand{\adfsh}{ADF$_{\text{Sh}}$}

\begin{document}

\title{Various Compressed Sensing Set-Ups Evaluated Against Shannon Sampling Under Constraint of Constant Illumination}

\author{Wouter~Van~den~Broek,$^1$\thanks{Email: vandenbroek@physik.hu-berlin.de} \ Bryan~W.~Reed,$^2$ Armand~B\'ech\'e,$^3$\\ Abner~Velazco,$^3$ Johan~Verbeeck,$^3$ and Christoph~T.~Koch.$^1$\\ \ \\ $^1$Institut f\"ur Physik, Humboldt Universit\"at zu Berlin,\\Newtonstr. 15, 12489 Berlin, Germany.\\ \ \\
$^2$Integrated Dynamic Electron Solutions, Inc., \\ 5653 Stoneridge Dr. Unit 117, Pleasanton CA 94588, US.\\ \ \\
$^3$EMAT, Antwerp University,\\Groenenborgerlaan 171, 2020 Antwerp, Belgium. }

\maketitle

\begin{abstract}
\textbf{%
Under the constraint of constant illumination, an information criterion is formulated for the Fisher information that compressed sensing measurements in optical and transmission electron microscopy contain about the underlying parameters. Since this approach requires prior knowledge of the signal's support in the sparse basis, we develop a heuristic quantity, the detective quantum efficiency (DQE), that tracks this information criterion well without this knowledge. It is shown that for the investigated choice of sensing matrices, and in the absence of read-out noise, i.e. with only Poisson noise present, compressed sensing does not raise the amount of Fisher information in the recordings above that of Shannon sampling.  Furthermore, enabled by the DQE's analytical tractability, the experimental designs are optimized by finding out the optimal fraction of on-pixels as a function of dose and read-out noise. Finally, we introduce a regularization and demonstrate, through simulations and experiment, that it yields reconstructions attaining minimum mean squared error at experimental settings predicted by the DQE as optimal.}
\end{abstract}


\section{Introduction}
\label{sec:intro}

Beam damage to the specimen is one of the most fundamental limits to the data quality in electron microscopy. In biological applications the acceptable dose is often below ten electrons per square \aa ngstr\"om, no matter if one images biological macromolecules  \cite{henderson1995,rez2003,baker2010} or records diffraction patterns of protein crystals  \cite{shi2013}. Although not as severe, beam sensitivity is an issue in materials science as well and researchers go to great lengths to limit it \cite{egerton2004,kaiser2011,jiang2012,egerton2015}. Furthermore, with the increased occurrence of soft and hard matter being interfaced into so-called hybrid materials  \cite{kirmse2016} biology's low upper bound for the electron dose now encroaches on the realm of materials physics too. A current trend in scanning transmission electron microscopy (STEM) seeks to limit electron exposure of the specimen by invoking compressed sensing \cite{romberg2008} (CS) in the recording process  \cite{stevens2014,beche2016,reed2017,donati2017}.

For applications with photon radiation, dose limitation is often a driving factor as well, for example when using potentially hazardous X-rays for computed tomography \cite{kubo2008}. Furthermore, limiting the recording time can be a valid goal in itself.

In CS, the signal $x$ is retrieved from the recordings $y$ that have been produced by the sensing matrix $A$ operating on $x$, i.e. $y = A x$. Recovery of $x$ from a surprisingly low number of measurements $y$ is possible if these measurements are incoherent  \cite{candes2007} to the signal and a sparsity constraint can be imposed. In general that involves expressing $x$ in a mathematical basis where it is sparse, for instance many photographs are sparse in a wavelet basis. For piecewise linear signals such an explicit decomposition can be omitted and $x$ can be retrieved instead by expressing that its total generalized variation (\tv) must be minimal; this approach is known as TV minimization \cite{candes2006}.

In order to make the measurements incoherent to the signal, sensing matrices conventionally have zero-mean independent and identically distributed (iid) random variables for entries. In the analysis of the respective error bounds, noise is often not considered, or assumed additive and/or bounded.

\subsection{Contribution of this paper}

In this paper it is acknowledged that in many experiments the total dose on the specimen is of greater importance than the number of measurements, and hence the performance of various CS set-ups is assessed under the constraint of constant total illumination. In other words, we investigate how to optimally make use of a given electron or photon budget.

Two fundamentally different and common set-ups are analyzed: the single-pixel set-up and annular dark field STEM (ADF-STEM).  These two instances respectively represent experiments where the illumination is caused by external sources outside of the scientist's control and those where they are in full command of the irradiation, between them covering most practical situations.  

The Fisher information that the measurements contain about the underlying parameters is evaluated through the A-optimality information criterion ($\crb$)  \cite{rao1945,cramer1946,lehmann2006,vandenbroek2009,vandenbroek2011}.  Furthermore, a heuristic quantity, the detective quantum efficiency (DQE), is developed.  With the aid of simulations it is shown that the $\dqe$ tracks the $\crb$ well.  Furthermore, a novel regularization is introduced and through simulations and experiment it is empirically demonstrated that the associated reconstructions attain minimum mean squared error (MSE) at experimental settings predicted by the DQE as optimal.

Having established the $\dqe$'s validity, its analytical tractability is used to show that for the investigated sensing matrices and in the absence of read-out noise, i.e. with only Poisson noise present, compressed sensing does not raise the amount of Fisher information in the recordings above that of a Shannon sampled signal.  This is reflected in the reconstruction results that yield a comparable MSE when both data sets are treated with the same algorithm to isolate the influence of the recording protocol.

Furthermore, the experimental designs are optimized, i.e. the fraction of on-pixels for best reconstruction quality is given as a function of particle dose, read-out noise and other experimental parameters.

\subsection{Relation to previous work}

The conventional zero-mean sensing matrices in, for instance, \cite{candes2006,candes2007} are not physically realizable for the particle-counting experiments investigated in this paper: the single-pixel camera and ADF-STEM. In a pivotal paper, Raginsky  \cite{raginsky2010} et al. have shown that in that context the use of a sensing matrix that preserves non-negativity and flux combined with the non-additive and dose dependent unbounded Poisson noise, has such a large and adverse effect on the derived error bounds that they do not even approach the conventional bounds in the limit of high dose and the associated relatively low noise.

CS has been analyzed from the perspective of Fisher information and the $\crb$ before \cite{babadi2009,miller2009,niazadeh2012,nielsen2012,pakrooh2013,pakrooh2016}.  These works, however, did not consider sensing matrices that preserve non-negativity and flux or the non-additive and dose dependent unbounded Poisson noise.  That doing so yields qualitatively different results is illustrated by the fact that the finding in \cite{nielsen2012} that ``the estimation accuracy [degrades] by at least the down-sample factor'' is reproduced in this work when just read-out noise is considered, but not when only Poisson noise is taken into account.

\subsection{Organization of the paper}

This paper is organized as follows. The image formation of ADF-STEM and the single-pixel camera is treated in Sec. \ref{sec:imafor}; the basics of compressed sensing pertaining to our problem are dealt with in Sec. \ref{sec:comsen}; in Sec. \ref{sec:staexp} Fisher information and the A-optimality information criterion are explained and the application of statistical experimental design to the compressed sensing set-ups is developed; the results are reported and discussed in Secs. \ref{sec:resdis} and \ref{sec:dis}; and in Sec. \ref{sec:con} the conclusions are drawn.

\section{Image formation}
\label{sec:imafor}

In this section the image formation for the single-pixel camera and ADF-STEM is established.

\subsection{Single-pixel camera}
\label{sec:sinpix}

\begin{figure}
  \center
  \includegraphics[width = 0.99\linewidth]{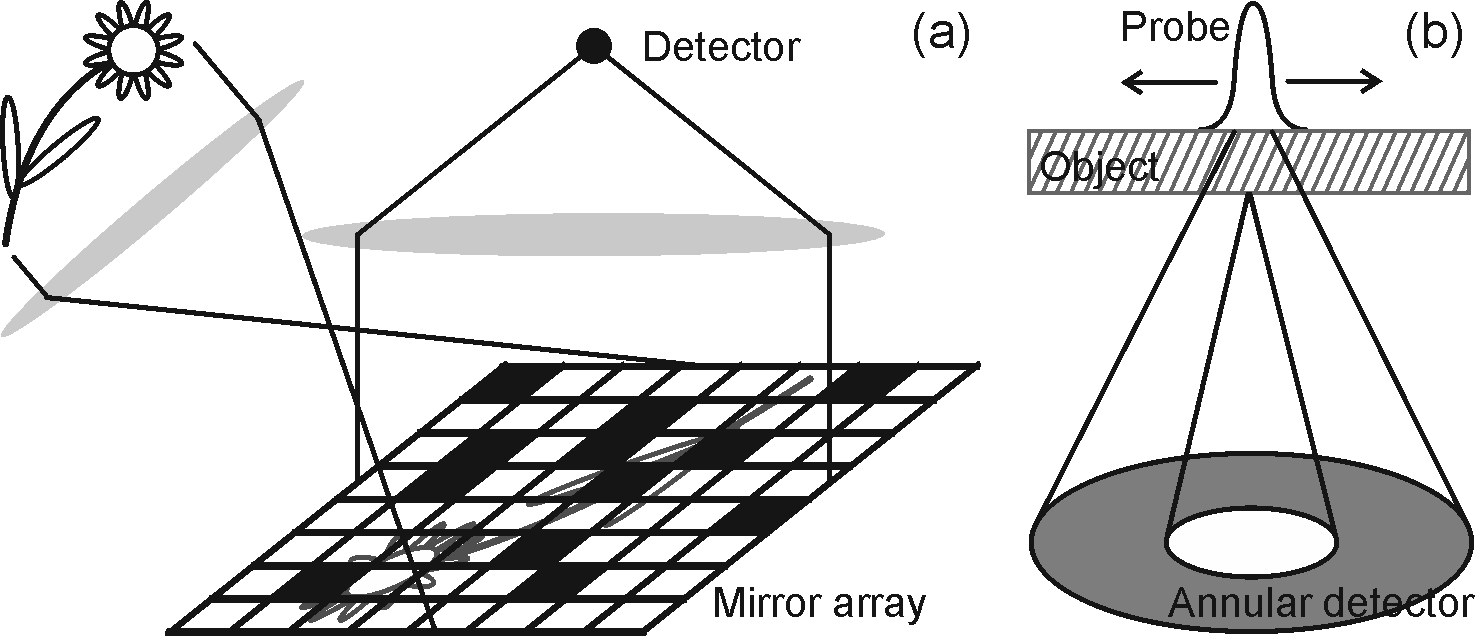}
  \caption{\label{fig:expsetup} \emph{(a)}: In the single-pixel camera the scene is projected onto an array of switchable mirrors, and the reflected light is focused onto the single-pixel detector. (Sketch after  \cite{baraniuk2007}) \emph{(b)}: In ADF-STEM an electron probe is scanned over the object, and electrons scattered to higher angles are integrated in an annular detector in the far-field.}
\end{figure}

A way of realizing a single-pixel camera \cite{duarte2008} is by projecting an image onto an array of switchable mirrors located in the image plane of an objective lens whose conjugate plane contains a photon detector on the optical axis; see Fig. \ref{fig:expsetup} for a sketch of the setup. Each of the mirrors in the array can be switched on or off, and fractional on-values can be obtained by tuning the pixels' on-time.

The linear image formation model is given as
\begin{eqnarray}
\label{eq:linimafor}
  \E(y) = I A x,
\end{eqnarray}
where the measurement vector $y$ has $M$ elements, the two-dimensional object is written in long-vector format as $x$ and has $N$ elements, the sensing matrix $A$ has dimensions $M \times N$ and $I \mu$ is the average number of photons reflecting off each of the object's pixels, with $\mu$ the average of $x$ and $\E$ denoting the expectation value.  Although not treated explicitly in this work, it is worthwhile noting that $I$ depends on $N$ through $I \propto 1 / N$, thus expressing that a finite number of photons must be budgeted over the $N$ pixels.

Two distinct choices for $A$ are investigated in this paper:
\begin{enumerate}
 \item[] \textbf{\sipidi}: each mirror in the array has a chance $p$ to provide the discrete value $v$ of $1$ (``on'') and a chance $1-p$ to provide $0$ (``off'');
 \item[] \textbf{\sipifr}: each mirror has a chance $p$ to provide a fractional value $v$ drawn from a uniform distribution on $[0,1]$, and a chance $1-p$ to provide $0$.
\end{enumerate}
This approach allows the value for $p$ to be optimized. Since the single-pixel camera is to be evaluated for a constant recording time, the elements of $A$ corresponding to on-pixels are set to $v/M$ and those corresponding to off-pixels to $0$, expressing that a constant expected total number of photons $NI\mu$ is reflecting off the object. Note that although the number of photons impinging on the sample is constant, the recorded number of photons scales with $p$.

The pixel-by-pixel raster scan, or Shannon sampling, is obtained with sensing matrix \sipidi \ when $p$ is set to $1/N$ and the matrix $A$ is diagonal with $M = N$. The subsequent reconstruction of the image under the application of a sparsity constraint then comes down to denoising.

The case of Gaussian distributed values in the sensing matrix is not considered in this paper.  It is trivial to adapt the conventional zero-mean binary and uniformly distributed values to the non-negative distributions considered here: shifting the distributions' mean accomplishes this without introducing distortions to the distributions.  A mere translation is not sufficient for the Gaussian distribution however, its non-boundedness implies additional clipping of remaining negative values \emph{and} too-high positive values.  The latter is necessary because a finite recording time sets an upper limit on the value of an on-pixel.  The analysis needed to separate the influence of such a distortion from the influence of the mere non-negativity falls outside of the scope of this work.

\subsection{ADF-STEM}
\label{sec:adfste}

For ADF-STEM the electron beam is condensed onto the specimen and scanned across it. For each beam position an annular detector in the far-field integrates the electrons scattered to its surface; see Fig. \ref{fig:expsetup}. Although analyzing  the resulting gray values in terms of the specimen's chemical composition \cite{hartel1996} is important, it is not the subject of this work and is hence not treated or attempted. Instead the signal $x$ is taken as the expected detector output for a certain beam position.

The image formation model is again given by (\ref{eq:linimafor}), and these choices of $A$ are investigated:
\begin{enumerate} 
  \item[] \textbf{\adfcs}: $A$'s elements are switched on with a probability $p$ and given a value of $1/(pN)$, off-pixels are set to $0$;
  \item[] \textbf{\adfsh}: $A$ is $N \times N$ with only the diagonal elements non-zero and equal to $M/N$.
\end{enumerate}
These two matrices allow comparison under equal dose in the recordings since they yield an expected total dose in $y$ of $MI\mu$, corresponding to an average of $I\mu$ and $IM\mu / N$ per individual measurement for sensing matrices \adfcs \ and \adfsh, respectively.

Note that in ADF-STEM the operator has direct control over the delivered dose through the beam intensity, dwell time and beam-blanking at will so that all electrons that scatter from the object and that could in principle be detected, actually are detected and contribute to the signal $y$; this recording mode is fundamentally different from the single-pixel case.

Sensing matrix \adfcs \ is analogous to the single-pixel set-up and can be experimentally realized through fast beam deflection as detailed in  \cite{beche2016,reed2017}; matrix \adfsh \ represents a classic raster scan, or Shannon scan, with a total intensity matched to the CS case.  A special case of \adfcs \ is obtained by equating $p$ to $1/N$ and corresponds to sampling the specimen in $M$ randomly selected points.  This is known as inpainting \cite{papafitsoros2014} and its treatment here is motivated by the interest it currently receives from various research groups \cite{stevens2014,beche2016}.

\subsection{Noise models}
\label{sec:noimod}

Since both imaging techniques are essentially particle counting experiments (photons or electrons) the measurements $y_i$ follow a Poisson probability density distribution
\begin{eqnarray}
\label{eq:poidis}
  \frac{ \(I A_i x\)^{y_i} }{ y_i! } \exp\( -I A_i x \)
\end{eqnarray}
with $A_i$ the $i^{\text{th}}$ row of $A$. This Poisson distribution has mean and variance $I A_i x$.

In the case the recording device is not perfect it could add read-out noise. To keep the problem analytically tractable the Poisson distribution is approximated with a normal distribution with mean and variance $I A_i x$, and the read-out noise is modeled as normal additive noise with zero mean and variance $c$. Since both contributions are independent their variances add, yielding
\begin{eqnarray}
\label{eq:nordis}
  \frac{1}{\sqrt{ 2 \pi \( I A_i x + c \) }} \exp\( \frac{-\(y_i - I A_i x\)^2}{2\( I A_i x + c \)} \)
\end{eqnarray}
as distribution for $y_i$.

The approximation of a Poisson distribution as a Gaussian is of great practical value in the theoretical analysis but needs some justification. In practice, the actual noise properties are rarely truly Poisson; the Poisson distribution is itself an approximation to the real system, and this provides some latitude to choose a more mathematically convenient approximate model.

For example, in ADF-STEM, the detection is indirect such that the signal produced by each electron is a random variable, and the aggregated signal yields a probability distribution that scales as Poisson but is not actually Poisson for several reasons. First, the distribution is continuous, not discrete. Second, the constant of proportionality between the signal and the variance is in general not unity. The effect of this calibration factor on the results in this article is nothing more than a trivial rescaling factor. Third, the readout noise also contributes, especially at low signal levels where it can be dominant. Thus, in the very regime where the Poisson distribution is most poorly approximated by a Gaussian, the dominant noise contribution is in fact Gaussian.

Since much of CS literature considers read-out noise only it is investigated in this paper too. The noise is assumed normal and additive with zero mean and variance $c$, so that $y_i$ is drawn from the distribution
\begin{eqnarray}
\label{eq:rdtdis}
  \frac{1}{\sqrt{ 2 \pi c }} \exp\( \frac{-\(y_i - I A_i x\)^2}{2 c } \).
\end{eqnarray}

\section{Compressed sensing}
\label{sec:comsen}

In compressed sensing (CS) under-determined or ill-posed problems are alleviated by regularization of the solution. In general a basis is defined in which the solution is expected to be sparse, and regularizing then means finding that solution with minimal $\ell_1$-norm in said basis.

An explicit decomposition can be avoided in the special case of piecewise linear objects $x$  with the aid of Hessian regularization \cite{scherzer1998,papafitsoros2014}.  To this end the total generalized variation ($\tv$) is defined as,
\begin{eqnarray}
\label{eq:tv}
  \tv(x) = \sum_k \absl s_k \absr, \text{ with } s = H x,
\end{eqnarray}
where $s_k$ is the Laplacian in $x_k$, calculated approximately by the operator $H$ which implements a convolution with the kernel
\begin{eqnarray}
 \begin{pmatrix}
  0& 1 & 0 \\
  1 & -4 & 1 \\
  0& 1 & 0
 \end{pmatrix}. \label{eq:lapKernel}
\end{eqnarray}
This approach is often used in inpainting \cite{papafitsoros2014}.

The signal $x$ can then be retrieved by solving the constraint optimization problem
\begin{eqnarray}
\label{eq:tvm}
  \min_x \tv(x) \ \text{ s.t. } \ IAx = y.
\end{eqnarray}
However, solving this system with exact compliance to the constraint leads to overfitting and a solution that cannot be considered very sparse anymore.  Instead, a constraint is stated that can be obeyed exactly without overfitting and can be complied with by minimizing an augmented Lagrangian \cite{nocedal1999} through an alternating direction scheme \cite{boyd2011,li2013,jiang2016}
\begin{eqnarray}
\label{eq:lnLElnL}
  \min_x \tv(x) \ \text{ s.t. } \ \ln L( y | x ) = \E \(  \ln L( y | x ) \).
\end{eqnarray}
$\ln L$ is the log-likelihood of the measurements $y$ conditional on the model parameters $x$.  More details and the equivalence to a convex optimization problem are shown in App. \ref{sec:ElnL}.  The encouraging results and close correspondence of this choice of regularization to the behavior predicted by the DQE are presented in this paper as empirical results.\footnote{The MATLAB implementation of this algorithm, sparseElnL, is made freely available (\url{https://github.com/woutervandenbroek/sparseElnL}).}

Although not needed for solving (\ref{eq:lnLElnL}), an explicit transformation
\begin{eqnarray}
\label{eq:Gs}
  x = G s,
\end{eqnarray}
is necessary for our analysis, changing the image formation model from (\ref{eq:linimafor}) into
\begin{eqnarray}
\label{eq:I0AGs}
  \E(y) = I A G s.
\end{eqnarray}

Since $s$ is computed through a convolution with kernel (\ref{eq:lapKernel}), the inverse operator $G$ can be obtained by transforming to Fourier space, inverting, zeroing the dc-component and transforming back to real space. As this resets the mean value to $0$, $s$ is extended by one element containing the mean, and the appropriate column is appended to $G$; similarly, a corresponding row is appended to $H$ as well.

\section{Statistical experimental design}
\label{sec:staexp}

In this section the Fisher information matrix, the A-optimality criterion ($\crb$) and the detective quantum efficiency (DQE) of the various measurement schemes are discussed.

\subsection{Fisher information matrix}
\label{sec:fiinma}

The Fisher information  \cite{frieden1998,lehmann2006} \fiin \ quantifies how much information the measurements $y$ contain about the unknown parameters $s$ that model it. In this section and in Sec. \ref{sec:crarao} the signal support is assumed known, i.e. only the non-zero elements of $s$ are retained. The Fisher information is defined as the negative expectation value of the curvature of the measurements' log-likelihood function $\ln L(y|s)$
\begin{eqnarray}
\label{eq:fiin}
  \fiin_{k,\ell}(s) = -\E \[ \frac{\partial^2 \ln L( y|s )}{\partial s_k \partial s_{\ell}} \].
\end{eqnarray}
If the measurements $y_i$ are drawn from the independent distributions $p_i(y_i|s)$, then
\begin{eqnarray}
  L( y|s ) = \prod_i  p_i( y_i | s ).
\end{eqnarray}

For the problems at hand, the distributions $p_i$ are given by the noise distributions in (\ref{eq:poidis}), (\ref{eq:nordis}) and (\ref{eq:rdtdis}). Working out (\ref{eq:fiin}) while taking (\ref{eq:Gs}) into account yields
\begin{eqnarray}
\label{eq:fipoi}
  \fiin_{k,\ell}(s) = I \sum_i \frac{[AG]_{i,k} [AG]_{i,\ell}}{ [AG]_i s }
\end{eqnarray}
for the Poisson distribution in (\ref{eq:poidis}), 
\begin{eqnarray}
\label{eq:finor}
\begin{split}
  \fiin_{k,\ell}(s) = & \ I \sum_i \frac{[AG]_{i,k} [AG]_{i,\ell}}{[AG]_i s + c/I} \bigg( 1 + \\ 
  \ & \frac{1}{2I\( [AG]_i s + c/I \)} \bigg)
\end{split}
\end{eqnarray}
for the case in (\ref{eq:nordis}) with Poisson and read-out noise, and
\begin{eqnarray}
\label{eq:firdt}
  \fiin_{k,\ell}(s) = \frac{I^2}{c} \sum_i [AG]_{i,k} [AG]_{i,\ell}
\end{eqnarray}
for read-out noise only, as described in (\ref{eq:rdtdis}).

In order to calculate $\fiin(s)$, all zero entries of $s$ are removed, as are the corresponding columns of $G$ and rows of $H$. It can be seen immediately from (\ref{eq:Gs}) that this does not affect the image formation.

\subsection{Information criterion}
\label{sec:crarao}

In statistical experimental design experiments are set up so as to maximize a measure of the information the recordings contain about the unknown parameters that model them.  This requires compression of the information matrix into a single number, a so-called information criterion, that then is optimized with respect to the experimental settings ($p$ in this paper).

For this work, the oft-used A-optimality is chosen.  It is the trace of the inverse of the information matrix, divided by the number of unknowns, $K$,
\begin{eqnarray}
\label{eq:crbtv}
  \crb = \tr \( \fiin(s)^{-1} \) / K.
\end{eqnarray}
Although various choices are possible, in \cite{cornell2002} it is noted that ``[A] design that is optimal for a given model using one [...] criteri[on] is usually near-optimal for the same model with respect to [...] other criteria.''

A-optimality has the added advantage that in case of unbiased estimators it serves as the lower bound on the mean squared error.   It is possible to attain this instance of the so-called Cram\'er Rao lower bound \cite{rao1945,cramer1946,lehmann2006} in practice; most notably  by a maximum likelihood estimation from a sufficient number of measurements \cite{lehmann2006}.  Nevertheless, one must keep in mind that foremost $\crb$ is a statement about the information contained in the measurements, independent of any estimation algorithm.  In view of the fact that the regularized estimates produced in CS are generally biased, the interpretation as a lower bound on the mean squared error is only of secondary importance in this context.\footnote{The MATLAB implementation of these calculations, sparseAOC, is made freely available (\url{https://github.com/woutervandenbroek/sparseAOC}).}

\subsection{Detective quantum efficiency}
\label{sec:dqe}

Rigorous as the $\crb$ is, it lacks generality in this context as it must be calculated for a particular sparsifying basis, example object $x$ and realization of $A$. The latter problem can be countered by averaging over multiple realizations of $A$, although this exacerbates an already demanding computation. Furthermore, it requires knowledge of the signal support and no analytical relation between the imaging system's settings and the reconstruction quality $\crb$ is provided.

In order to gain insight, the detective quantum efficiencies (DQE) of the various CS setups are introduced.  While conventionally the DQE characterizes noise properties of imaging devices \cite{rose1946,niermann2012}, it is slightly generalized here in order to characterize the recording set-up as a whole,
\begin{eqnarray}
  \dqe = \alpha \frac{\snrout^2}{\snrin^2}. \label{eq:dqe}
\end{eqnarray}
$\snrout$ and $\snrin$ are the signal-to-noise ratios of the recorded signal and the incoming signal, respectively.  The prefactor $\alpha$ reflects the different number of measurements in both cases, hence $\alpha = M/N$ for sensing matrices \sipidi, \sipifr \ and \adfcs, and $\alpha = 1$ for sensing matrix \adfsh.  More details are provided in App. \ref{sec:derdqe}.

The SNR is defined as the ratio of the standard deviation of the signal to the standard deviation of the noise.  For $\snrout$ this is calculated as the recorded signal $y$, while for $\snrin$ we use the best possible hypothetical reference signal allowed by Poisson noise given the number of available particles.  In case of the single-pixel camera this constitutes the signal that would be obtained with an ideal detector in lieu of each of the mirrors in the image plane, while for ADF-STEM it means spreading out the available dose over all pixels in the image; the respective expressions are (\ref{eq:snrinsp}) and (\ref{eq:snrinADF}).

Since the DQE is calculated from $y$ instead of $s$ knowledge of the support does not come into play.  Although more heuristic than the $\crb$, the DQE is more general as $s$ enters its expression through the average $\mu$ of $x$ only, and is independent of the particular realization of $A$ and only depends on the variance of $A$'s elements.

The results are summarized in Table \ref{tab:dqes}, where the new variables
\begin{eqnarray}
  r = \frac{M}{N} \text{, } \gamma = \frac{c N}{I \mu} \text{ and } \gamma' = \frac{c}{I \mu}, \label{eq:varssipi}
\end{eqnarray}
have been used. $r$ is the reduction, and $\gamma$ and $\gamma'$ the normalized variances of the read-out noise.

Note how the last column of the first two rows of Table \ref{tab:dqes} reproduces the $N/M$ dependency of the $\crb$ presented in Eq. 21 in \cite{nielsen2012} for read-out noise without Poisson statistics.

\begin{table}
\renewcommand{\arraystretch}{1.5}
\caption{DQEs for Various Sensing Matrices and Noise Models. }
\label{tab:dqes}
\center
\begin{normalsize}
\begin{tabular}{c | c c c }
  \hline
	\hline
	{\footnotesize Sensing} & \multicolumn{3}{c}{{\footnotesize Noise model}}\\
  {\footnotesize matrix}  & {\footnotesize Poisson} & {\footnotesize Poisson + read-out} & {\footnotesize read-out} \\
  \hline
   {\footnotesize \sipidi}  & $\frac{1-p}{N}$ & $\frac{p(1-p)}{pN + r\gamma}$ & $\frac{p(1-p)}{r \gamma}$ \\
   {\footnotesize \sipifr}  & $\frac{2/3 - p/2}{N}$ & $\frac{p(1/3-p/4)}{pN/2 + r\gamma}$ & $\frac{p(1/3-p/4)}{r\gamma}$ \\
   {\footnotesize \adfcs}  & $\frac{1-p}{pN}$ & $\frac{1-p}{pN}\frac{1}{1 + \gamma'}$ & $\frac{1-p}{pN}\frac{1}{\gamma'}$ \\
   {\footnotesize \adfsh}  & $1$ & $\frac{1}{1+\gamma'/r}$ & $\frac{r}{\gamma'}$ \\
	\hline
	\hline
\end{tabular}
\end{normalsize}
\end{table}

\section{Results}
\label{sec:resdis}

With the aid of simulations a good agreement between $\crb$ and $\dqe$ is demonstrated.  Furthermore, reconstructions from a variety of simulated single-pixel set-ups and from an ADF-STEM experiment show close correspondence between MSE and $\dqe$.  This justifies the use of the $\dqe$ to derive optimal experimental settings and to determine when a CS set-up is preferable over a denoised Shannon scan.

\subsection{Agreement between DQE and $\crb$}
\label{sec:annuev}

The agreement between DQE and $\crb$ for single-pixel camera and ADF-STEM is evaluated under the assumption of three different noise models: Poisson noise only, Poisson noise and read-out noise and read-out noise only.

\subsubsection{Single-pixel}
\label{sec:crbdqesipi}

Two distinct single-pixel set-ups are investigated: \sipidi \ where the mirrors in the array take on discrete values of either $0$ or $1$, and \sipifr \ where they take a fractional value between $0$ and $1$.

\begin{figure}
  \center
  \includegraphics[width = 0.99\linewidth]{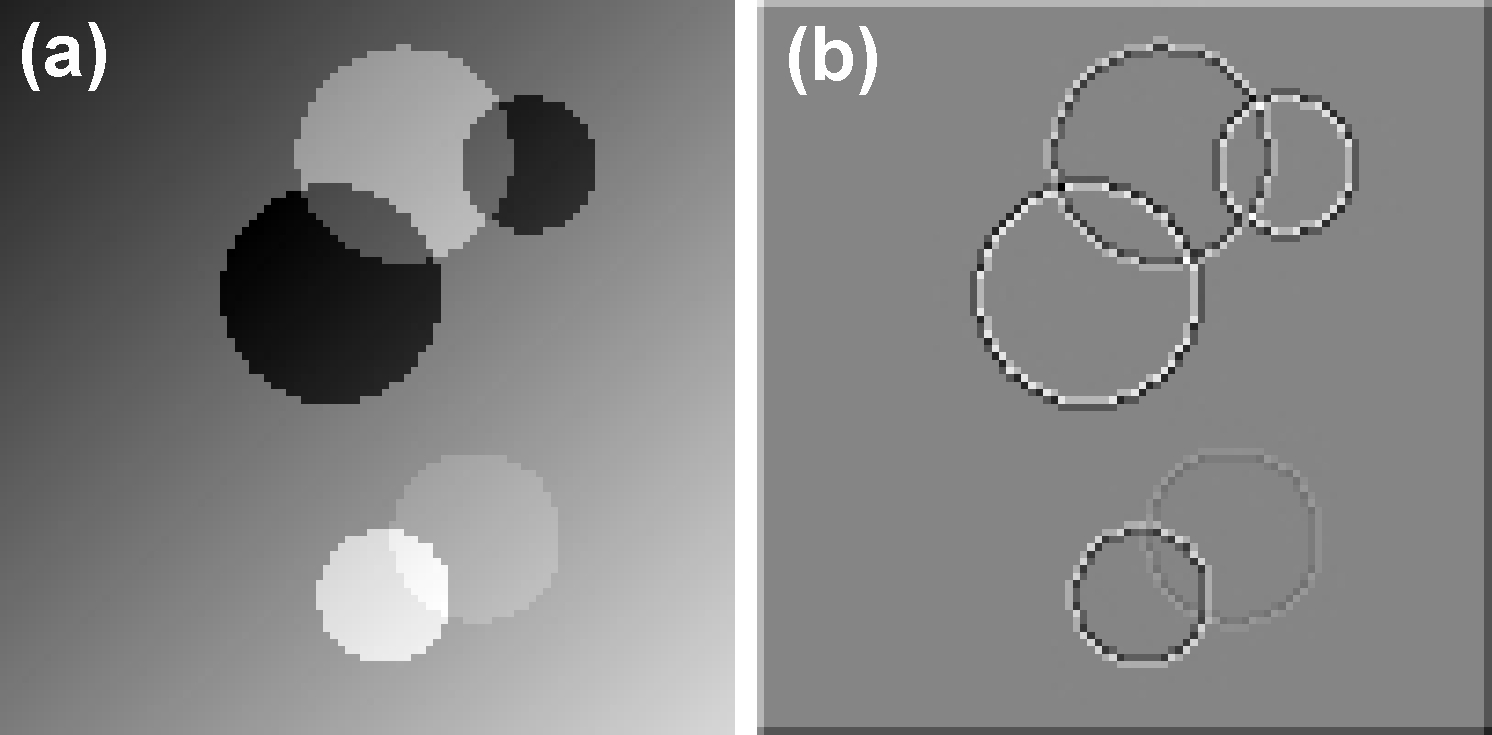}
  \caption{\label{fig:shepplogan} \emph{(a)}: The $100 \times 100$ Ramp-Discs phantom with gray values in the interval $[0.1, 1.0]$.  \emph{(b)}: The absolute values of the Laplacian.}
\end{figure}

The test sample $x$ is the $100 \times 100$ Ramp-Discs phantom, displayed in Fig. \ref{fig:shepplogan}, yielding $N = 10000$. The intensities lie in the interval $[0.1, 1]$ to better mimic realistic experimental conditions; $\mu = 0.53$ and $\sigma = 0.18$. The sparse vector $s$ has approximately $N/10$ non-zero elements and $M$ is set to double that value, i.e. $2000$, to arrive at a reduction $r$ of $20$\%. The relative read-out noise $\gamma$ is set to $625$ and $I = 7.5 \times 10^5$, ensuring that $\popt=0.10$ and that a CS-measurement with sensing matrix \sipidi, $p = 0.01$ and combined Poisson and read-out noise has a $\snrout^2$ of $10$ as calculated with (\ref{eq:snrout}).

\begin{figure}
  \center
  \includegraphics[width = 0.99\linewidth]{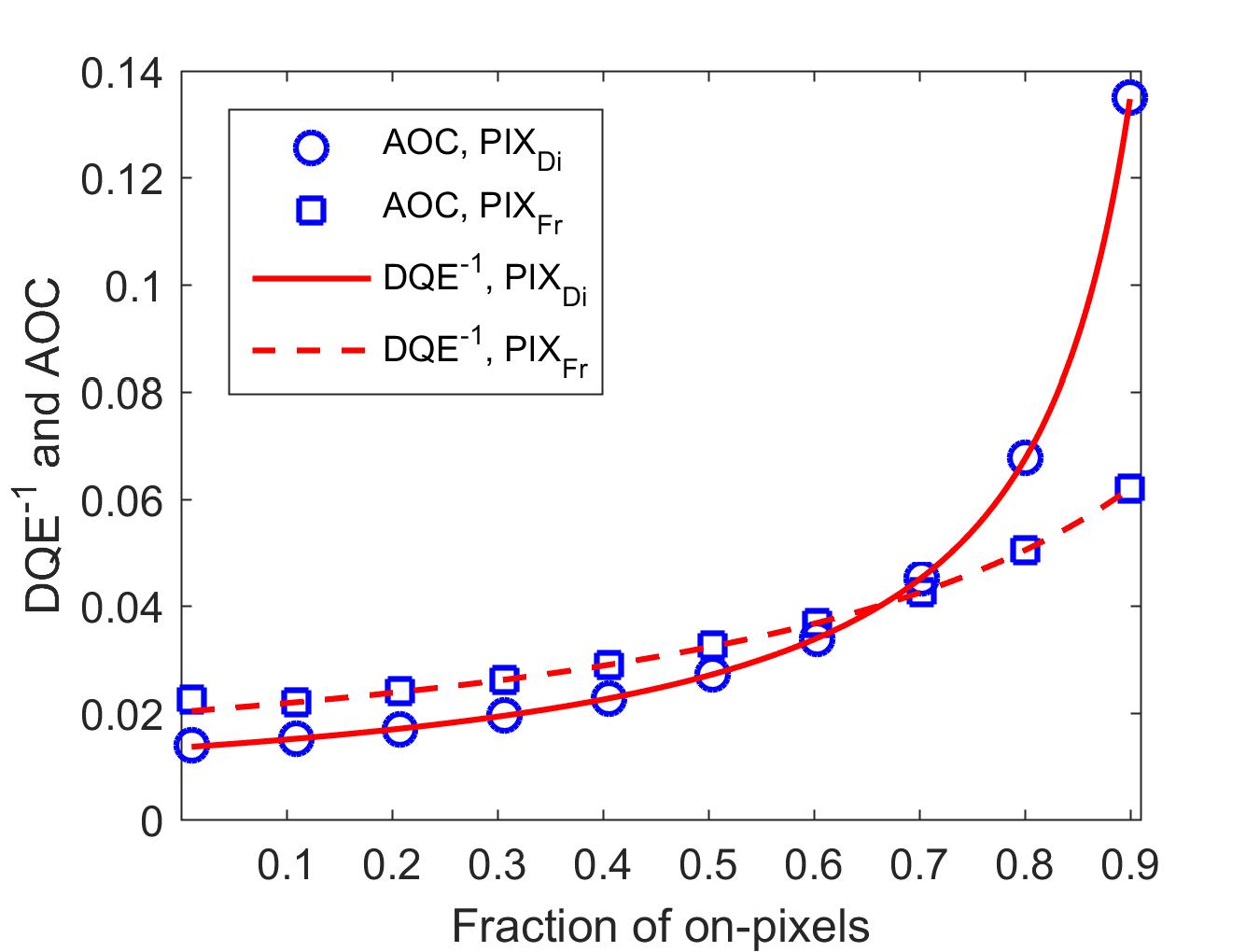}
  \caption{\label{fig:sipipois} $\dqe^{-1}$ and $\crb$ vs. the fraction of on-pixels for the single-pixel set-up for Poisson noise only.}
\end{figure}

The results for Poisson noise only are depicted in Fig. \ref{fig:sipipois}. Predictions by $\dqe^{-1}$ agree well with the $\crb$, except for very low $p$ for sensing matrix \sipifr.

\begin{figure}
  \center
  \includegraphics[width = 0.99\linewidth]{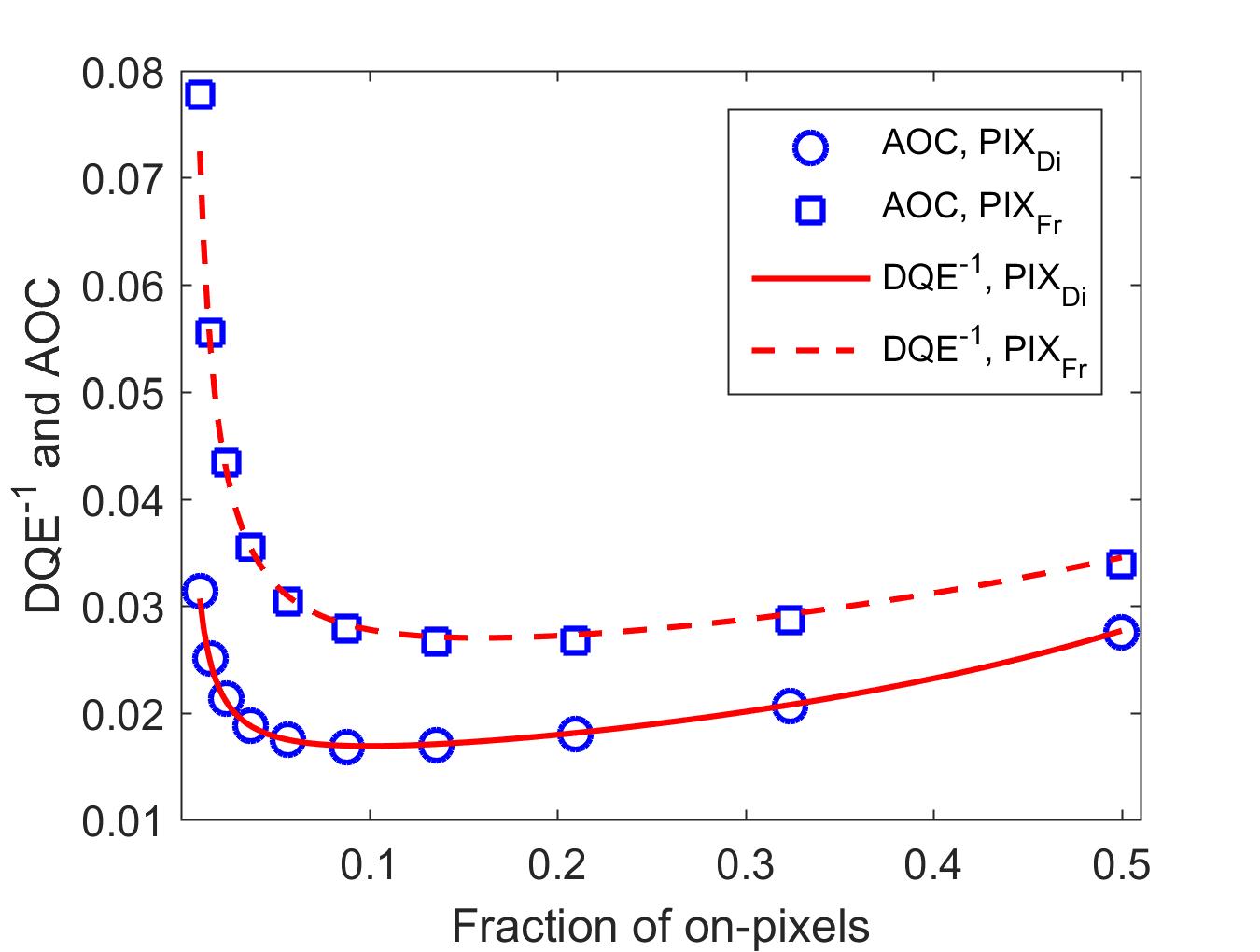}
  \caption{\label{fig:sipipoisrdt}  $\dqe^{-1}$ and $\crb$ vs. the fraction of on-pixels for the single-pixel set-up for simultaneous Poisson and read-out noise. The values for $p$ are logarithmically spaced to better sample the region for low $p$.}
\end{figure}

For these simulation settings the optimal value $\popt$ for the case of simultaneous Poisson and read-out noise is $0.10$ for \sipidi \ and $0.16$ for \sipifr \ as given by (\ref{eq:sipioptpoisrdt}) and (\ref{eq:sipioptpoisrdt2}) respectively. In Fig. \ref{fig:sipipoisrdt} it is shown how $\crb$ and $\dqe^{-1}$ coincide for a wide range of $p$.

\begin{figure}
  \center
  \includegraphics[width = 0.99\linewidth]{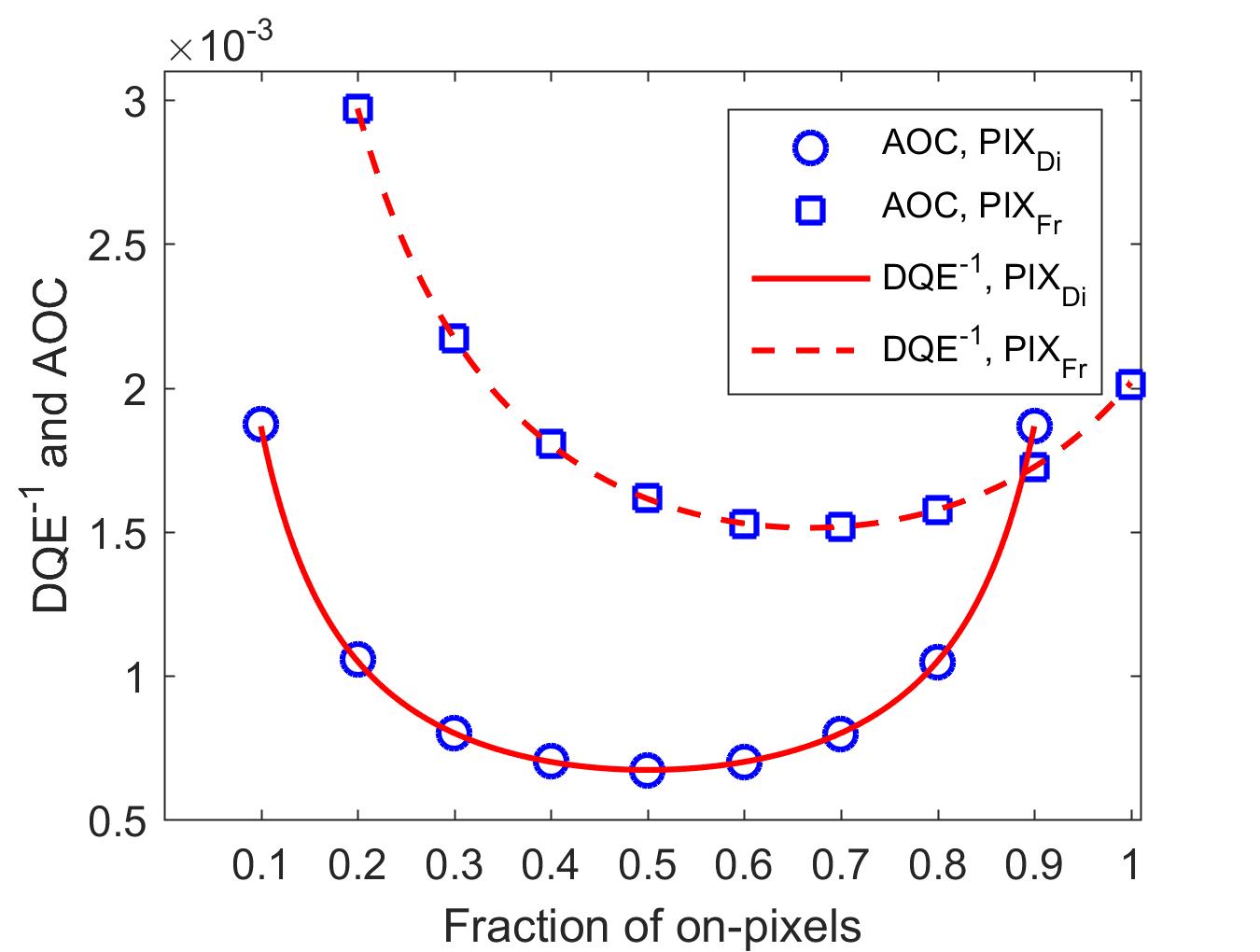}
  \caption{\label{fig:sipirdt}  $\dqe^{-1}$ and $\crb$ vs. the fraction of on-pixels for the single-pixel set-up for just read-out noise. }
\end{figure}

The results for read-out noise only are depicted in Fig. \ref{fig:sipirdt}.  For both sensing matrices, \sipidi \ and \sipifr, $\dqe^{-1}$ and $\crb$ agree well; yielding an optimum at $0.50$ and $0.67$, respectively.

\subsubsection{ADF-STEM}
\label{sec:crbdqeadf}

The CS implementation of ADF-STEM, sensing matrix \adfcs, is tested for all three noise models.

Simulations are carried out on the $100 \times 100$ Ramp-Discs phantom displayed in Fig. \ref{fig:shepplogan}. The simulation parameters are identical to those in Sec. \ref{sec:sipicare}, except that $\gamma'$ has been set to $0.5$ and $I = 2.5 \times 10^4$ so that the \adfcs -measurement with $p = 0.01$ and combined Poisson and read-out noise has a $\snrout^2$ of $10$.

\begin{figure}
  \center
  \includegraphics[width = 0.99\linewidth]{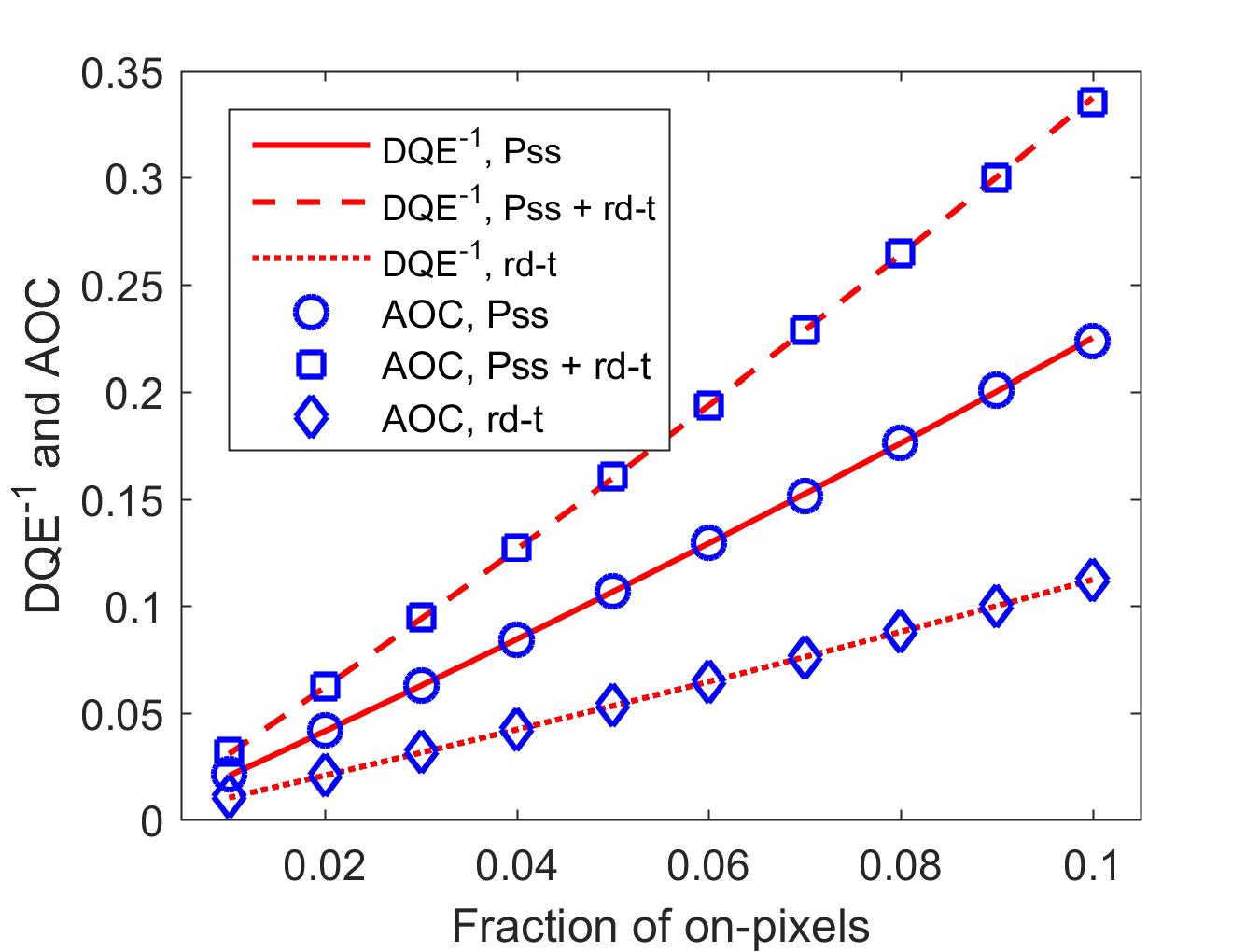}
  \caption{\label{fig:adfstemresults}  $\dqe^{-1}$ and $\crb$ vs. the fraction of on-pixels for sensing matrix \adfcs. All noise models: Pss stands for Poisson noise, rd-t for read-out noise.}
\end{figure}

The values for $\dqe^{-1}$ are compared to $\crb$ in Fig. \ref{fig:adfstemresults}.  A least absolute differences fit is used to scale both curves to each other, and an excellent agreement can be observed.

\subsection{Agreement between DQE and MSE}
\label{sec:exre}

In this section Monte Carlo simulations and experiments are presented to illustrate that the settings predicted by the DQE as optimal do yield reconstructions with minimal mean squared error (MSE) when regularized with the $\ln L = \E(\ln L)$ constraint in Sec. \ref{sec:comsen}.

\subsubsection{Monte Carlo simulations}
\label{sec:mocasi}

Single-pixel and ADF-STEM CS-measurements of the Ramp-Discs phantom were simulated with the same settings as in Sec. \ref{sec:crbdqesipi}.  Poisson noise and additive normal noise were added to the recordings as needed. For each value of $p$ ten measurements were simulated with different realizations of sensing matrix and noise.  As starting guess 
\begin{eqnarray}
  \frac{ A^T y }{ I \langle A \rangle^2 N M } \label{eq:stague}
\end{eqnarray}
with a slight perturbation was used, where $\langle A \rangle$ denotes the average of all elements of $A$.

The MSEs were calculated with respect to the original phantoms and their averages plotted as a function of $p$ along with the sample standard deviation.  A first order least squares fit was used to match $\dqe^{-1}$ to MSE.  That, contrary to Sec. \ref{sec:annuev}, a mere scaling does not suffice for a good agreement is an indication of the general biasedness of regularized estimators.

\begin{figure}
  \center
  \includegraphics[width = 0.99\linewidth]{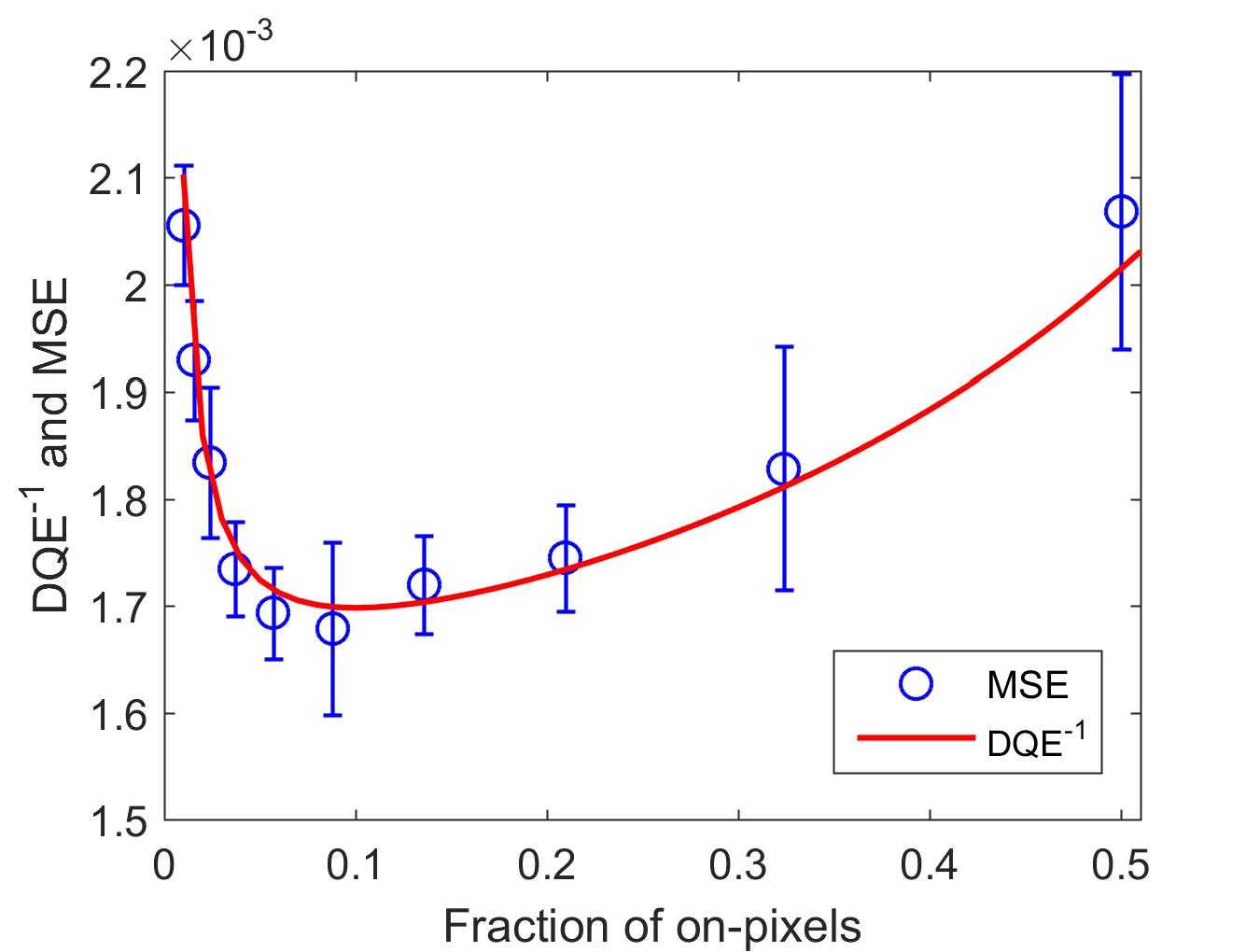}
  \caption{\label{fig:sM1-nM2_sim} $\dqe^{-1}$ and $\mse$ vs. the fraction of on-pixels for the single-pixel camera with discrete on-values (\sipidi) with Poisson noise and read-out noise. The values of $p$ are spaced logarithmically to sample the region around the optimum better. Compare to the results in Fig. \ref{fig:sipipoisrdt}.}
\end{figure}

The results for the single-pixel camera with discrete mirror values (\sipidi) in the presence of Poisson noise and read-out noise are presented in Fig. \ref{fig:sM1-nM2_sim}. The optimization ran for $50$ iterations, with $10$ subiterations for step 1; see App. \ref{sec:ElnL}. The MSEs show the same characteristic optimum for $p = 0.10$ as above. The $\dqe^{-1}$ matches the MSE well and predicts the minimum.

\begin{figure}
  \center
  \includegraphics[width = 0.99\linewidth]{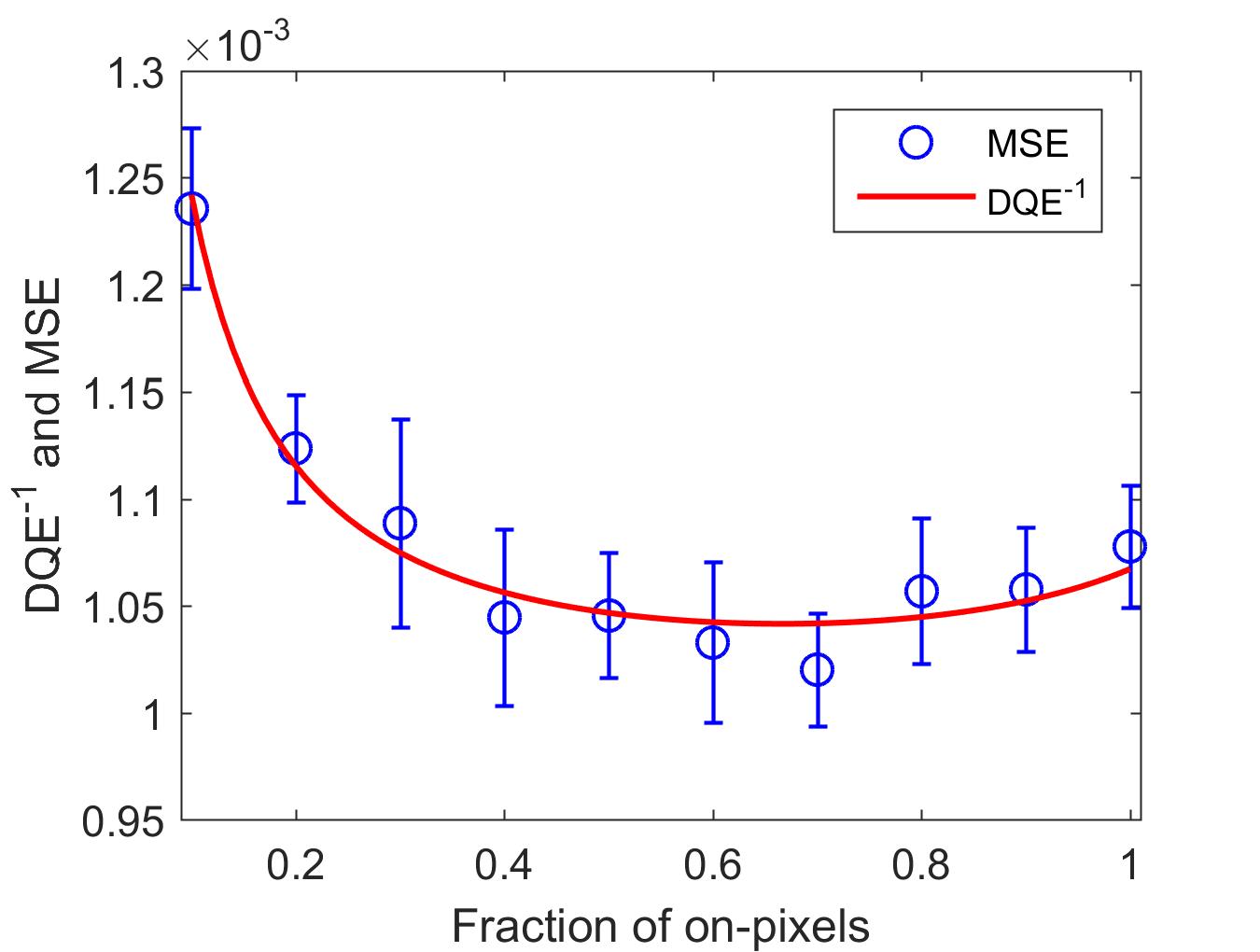}
  \caption{\label{fig:sM6-nM3_sim} $\dqe^{-1}$ and $\mse$ vs. the fraction of on-pixels for the single-pixel camera with fractional on-values (\sipifr) for read-out noise only. Compare to the results in Fig. \ref{fig:sipirdt}.}
\end{figure}

The results for the single-pixel set-up with fractional mirror values (\sipifr) in the presence of just read-out noise are presented in Fig. \ref{fig:sM6-nM3_sim}. The optimization ran for $100$ iterations, with $10$ subiterations for step 1. The MSEs show the same characteristic optimum at about $0.67$ as above. The $\dqe^{-1} $ matches the MSE well and can hence be used to predict the optimal experimental settings.

\begin{figure}
  \center
  \includegraphics[width = 0.99\linewidth]{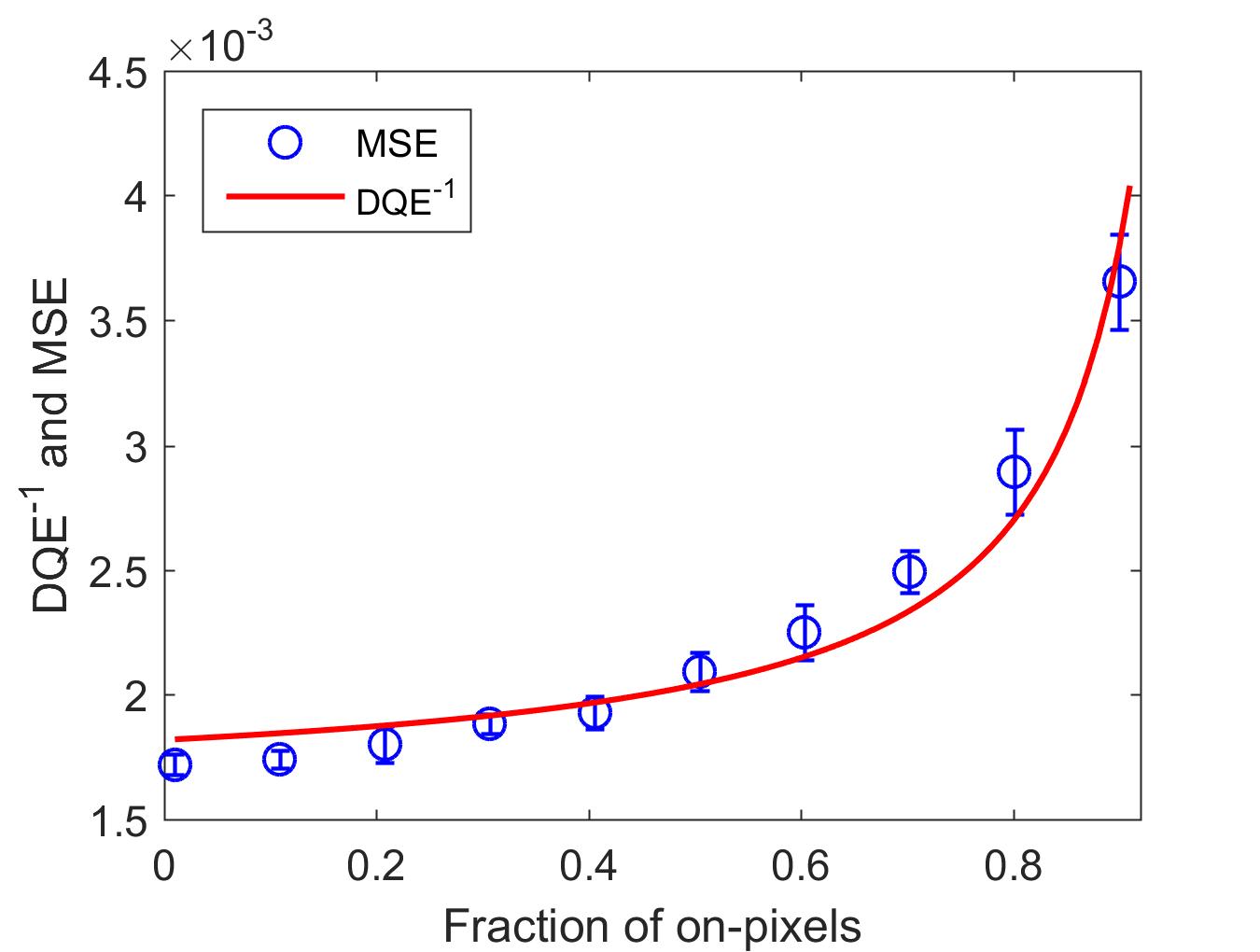}
  \caption{\label{fig:sipi_poiss_sim} $\dqe^{-1}$ and $\mse$ vs. the fraction of on-pixels for the single-pixel set-up with discrete on-values (\sipidi) for Poisson noise only. Compare to the results in Fig. \ref{fig:sipipois}.}
\end{figure}

In Fig. \ref{fig:sipi_poiss_sim} results for the single-pixel camera with discrete on-values (\sipidi) in the presence of just Poisson noise are presented.  The optimization ran for $50$ iterations, with $10$ subiterations for step 1. Despite systematic deviations, the $\dqe^{-1}$ shows the same characteristic monotonic increase as the MSE and hence predicts the optimal settings well.

Ten values for $p$ were chosen, equidistantly spaced from $0.01$ to $0.90$.  To avoid purely numerical difficulties, the variable $\gamma$ was set to $0.05$ so that according to (\ref{eq:sipioptpoisrdt}) the optimal value for $p$ equals $0.001$, i.e. ten times smaller than the lowest value encountered.  This makes the ratio between the terms $c$ and $IAx$ in constraint (\ref{eq:clnL}) of the order of $10^{-4}$ for $p = 0.01$; for higher $p$ the ratio is even smaller.
 
\subsubsection{Experimental ADF-STEM}
\label{sec:expadfstem}

\begin{figure}
  \center
  \includegraphics[width = 0.99\linewidth]{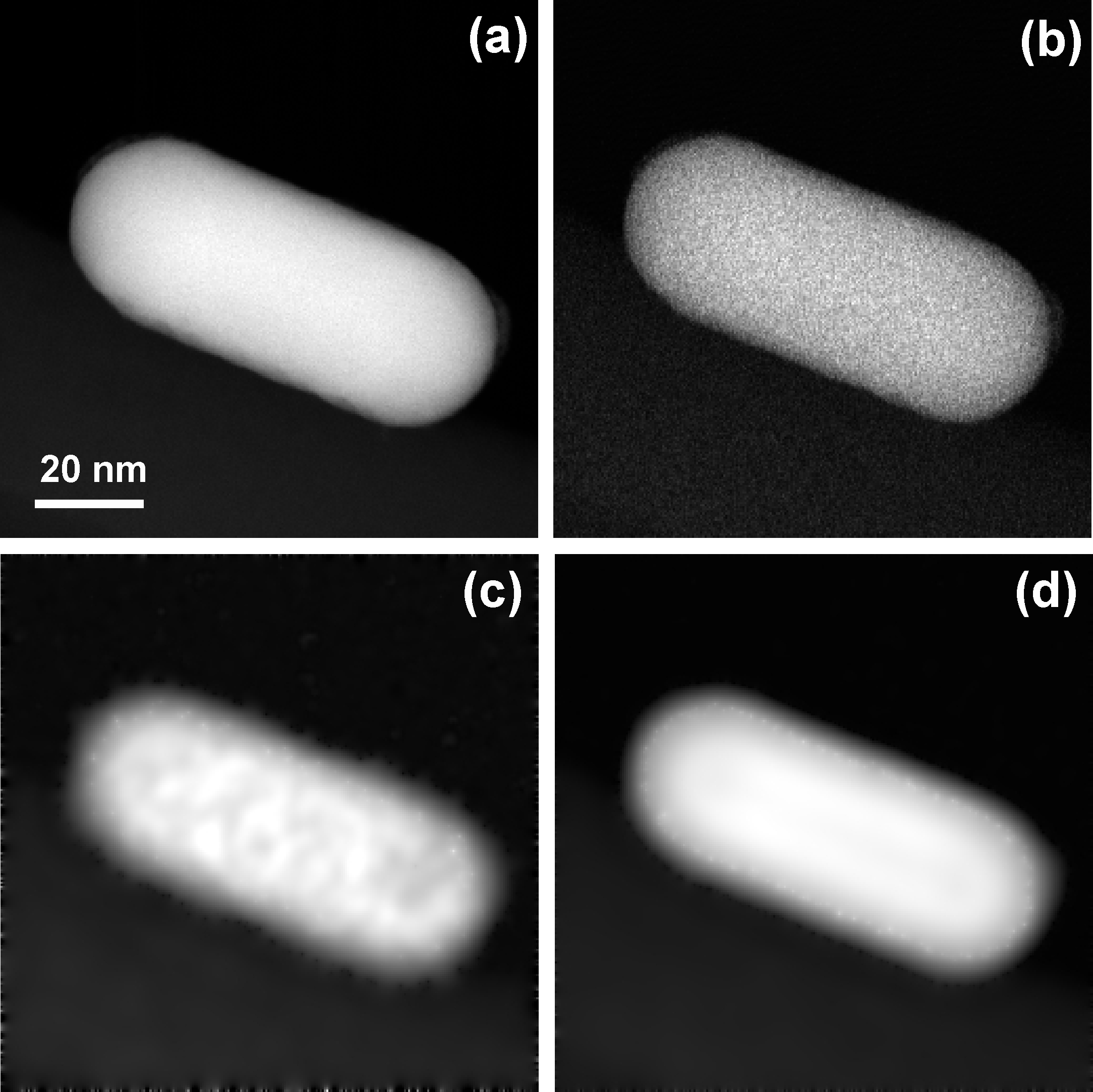}
  \caption{\label{fig:adfstem1} Experimental ADF-STEM data. \emph{(a)} Ground truth image $x_0$. \emph{(b)} Image $x_1$. \emph{(c)} Reconstruction from $M = N/10$ measurements, with $p = 1/N$. \emph{(d)} Reconstruction from $M = N$ measurements with $p = 1/N$, i.e. a denoising.}
\end{figure}

A gold nanorod was imaged in a FEI Titan$^3$ microscope operating in STEM mode at 300 kV.  An image $x_0$ with $256 \times 256$ pixels was raster scanned with an intensity $I_0 = 3.9\times 10^3$. After application of a $3 \times 3$ median filter, it served as ground truth to compute the MSE of the reconstructions; see Fig. \ref{fig:adfstem1}.a. Then, a $256 \times 256$ image $x_1$ was raster scanned with an intensity of $I _1 = 1.1 \times 10^2$; see Fig. \ref{fig:adfstem1}.b. At a standard deviation $\sqrt{c}$ of $0.32$~$\elctr$ and $0.43$~$\elctr$ for $x_0$ and $x_1$, respectively, the read-out noise was small, and this experiment exhibits almost pure Poisson noise.  

Ten sensing matrices of type \adfcs \ were constructed with $p = 1/N$ and $M$ ranging from $N/10$ to $N$ in steps of $N/10$.  This corresponds to an inpainting set-up with a fraction $r$ of $M/N$ pixels available.  From $x_1$ and these ten matrices, ten measurements $y_1$ were synthesized.

The DQE for this set-up is calculated as laid out in App. \ref{sec:derdqe}.  However, since in this case the measurements are not taken at equal dose, (\ref{eq:snrinsp}) must be used for $\snrin^2$ instead of (\ref{eq:snrinADF}); resulting in $\dqe = M / N$.

\begin{figure}
  \center
  \includegraphics[width = 0.99\linewidth]{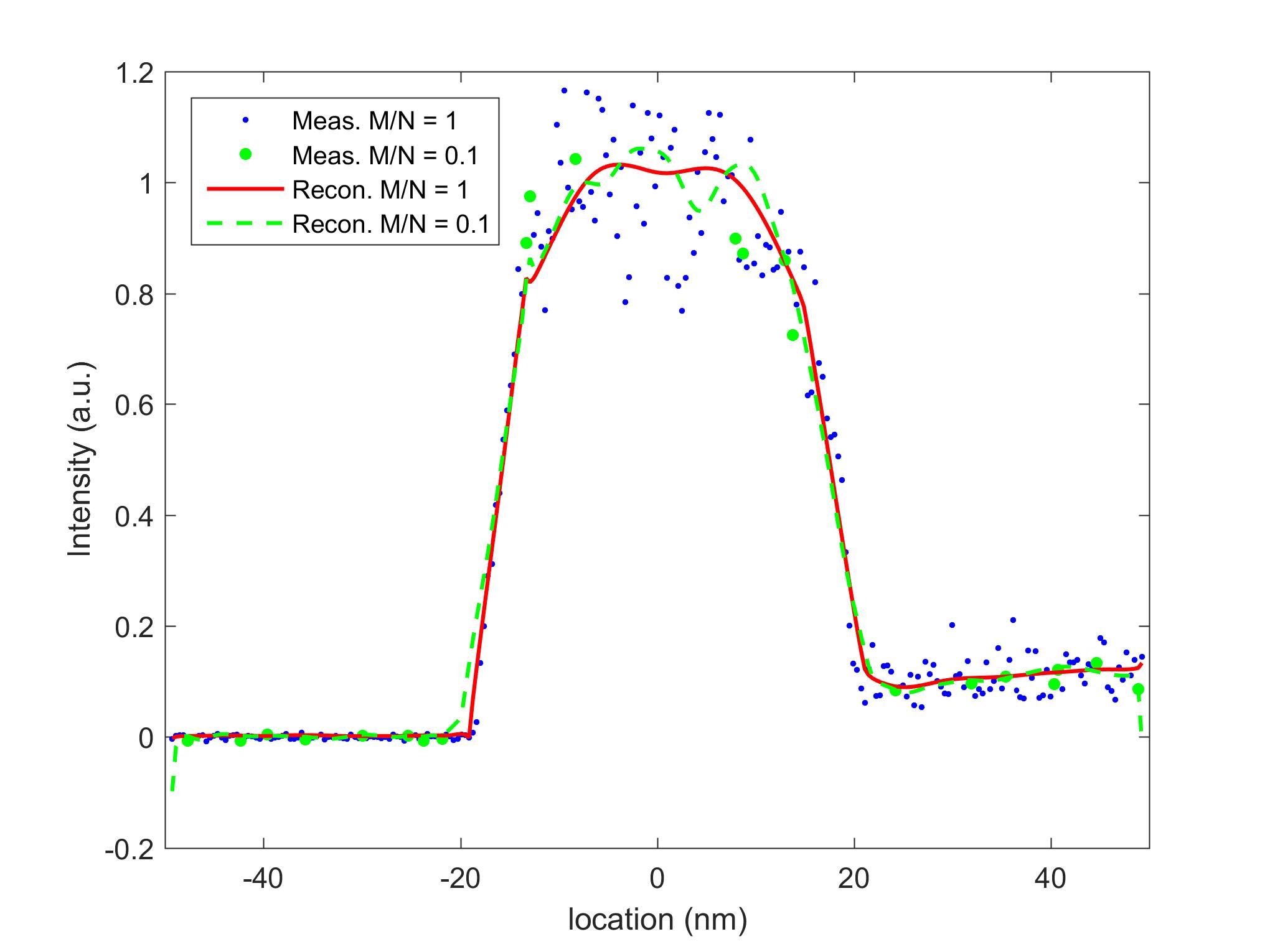}
  \caption{\label{fig:adfstem2Profiles}  Profiles and measurements for the central vertical line of the reconstructions in Fig. \ref{fig:adfstem1}, for $M/N = 0.1$ and $M/N = 1$. }
\end{figure}

Again constraint (\ref{eq:clnL}) was used for the reconstructions.  The starting guesses were set according to (\ref{eq:stague}).  The reconstructions ran for $200$ iterations, with $10$ subiterations for step 1.  In Figs. \ref{fig:adfstem1}.c and \ref{fig:adfstem1}.d results are shown for $M = N/10$ and $M = N$.   Profiles over the central vertical line of the reconstructions are contrasted with the measurements in Fig. \ref{fig:adfstem2Profiles} to allow visual assessment of the reconstruction quality .

\begin{figure}
  \center
  \includegraphics[width = 0.99\linewidth]{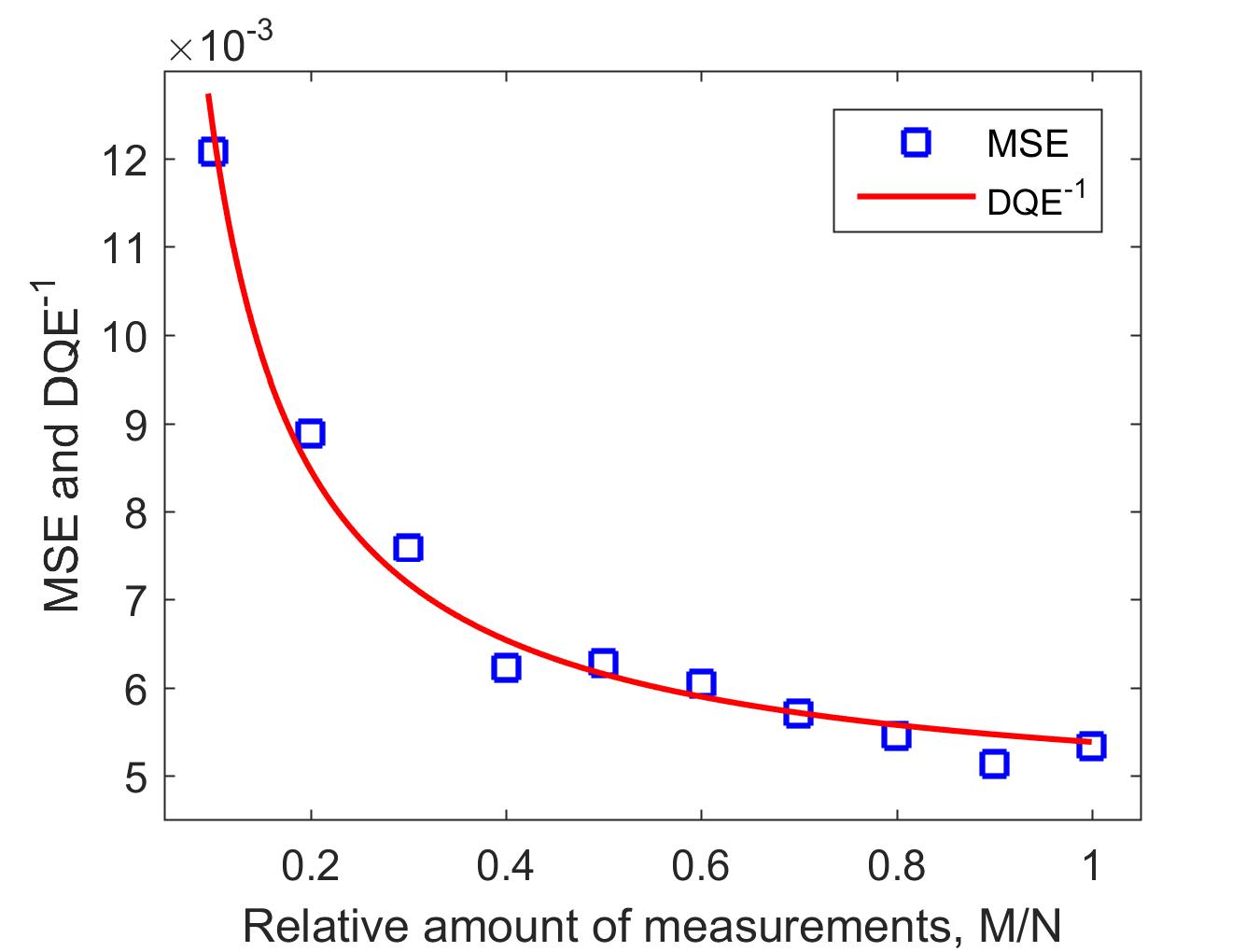}
  \caption{\label{fig:adfstem2}  After a linear transformation, the $\dqe^{-1}$s match well to the MSEs of the reconstructions from the experimental ADF-STEM data. }
\end{figure}

For calculation of the MSEs, the reconstructions were registered to $x_0$ and the intensities were matched by a linear least squares fit to $x_0$.  The resulting MSEs are depicted in Fig. \ref{fig:adfstem2} where they are shown to match well with $\dqe^{-1}$ after a linear transformation has been applied to the latter.

\subsection{Main results: DQE analysis}
\label{sec:dqeres}

In this section optimal values $\popt$ for $p$ are derived from the DQE; the results are summarized in Table \ref{tab:popt}. Furthermore, the $\dqe$'s analytical tractability is deployed to investigate if the denoising of a Shannon scan is preferable over a full CS set-up, and under what circumstances this might be the case.

\begin{table}
\renewcommand{\arraystretch}{1.5}
\caption{Optimal Values for $p$ for Various Sensing Matrices and Noise Models.  See also (\ref{eq:sipioptpoisrdta}) and (\ref{eq:sipioptpoisrdta2}).}
\label{tab:popt}
\center
\begin{footnotesize}
\begin{tabular}{c | c c c }
  \hline
	\hline
	{\footnotesize Sensing} & \multicolumn{3}{c}{{\footnotesize Noise model}}\\
  {\footnotesize matrix}  & {\footnotesize Poisson} & {\footnotesize Poisson + read-out} & {\footnotesize read-out} \\
  \hline
   {\footnotesize \sipidi}  & $1/N$ & $\sqrt{r\gamma/N}$ & $1/2$ \\
   {\footnotesize \sipifr}  & $1/N$ & $\frac{4}{\sqrt{6}}\sqrt{r\gamma/N}$ & $2/3$ \\
   {\footnotesize \adfcs}  & $1/N$ & $1/N$ & $1/N$ \\
   {\footnotesize \adfsh}  & n.a. & n.a. & n.a. \\
	\hline
	\hline
\end{tabular}
\end{footnotesize}
\end{table}

\subsubsection{Single-pixel camera}
\label{sec:sipicare}

In the presence of just Poisson noise the DQE decreases monotonically with $p$, thus suggesting that the optimal value is the minimum $1/N$, which defines an inpainting set-up.  In the case of Poisson noise and read-out noise, the optimal value is obtained by equating the derivative of $\dqe$ with respect to $p$ to zero, yielding
\begin{eqnarray}
  \popt & = & \frac{r \gamma}{N} \( \sqrt{1 + \frac{N}{r\gamma}} - 1 \), \label{eq:sipioptpoisrdta} \\
        & \simeq & \sqrt{ \frac{r\gamma}{N} }, \quad \text{for} \ \sqrt{ \frac{r\gamma}{N} } \ll 1, \label{eq:sipioptpoisrdt}
\end{eqnarray}
for sensing matrix \sipidi. For sensing matrix \sipifr \ it holds that,
\begin{eqnarray}
  \popt & = & 2\frac{r \gamma}{N} \( \sqrt{1 + \frac{2}{3} \frac{N}{r\gamma}} - 1 \), \label{eq:sipioptpoisrdta2} \\
        & \simeq & \frac{4}{\sqrt{6}}\sqrt{ \frac{r\gamma}{N} }, \quad \text{for} \ \sqrt{ \frac{r\gamma}{N} } \ll 1. \label{eq:sipioptpoisrdt2}
\end{eqnarray}
For just read-out noise an optimal value of $1/2$ is reached for \sipidi, and $2/3$ for \sipifr. See Table \ref{tab:popt}

The DQE for the Shannon case is derived by setting $p = 1/N$ and $r = 1$ in the expression for \sipidi \ with simultaneous Poisson and read-out noise (Table \ref{tab:dqes}, first row, second column), yielding
\begin{eqnarray}
  \dqe_{\text{Sh}} =  \frac{1}{N(1 + \gamma)}.
\end{eqnarray}

The CS set-up is treated for low read-out noise by evaluating the $\dqe$ in the respective optimal values for $p$ (\ref{eq:sipioptpoisrdt}):
\begin{eqnarray}
  \dqe_{\text{CS}} \simeq \frac{1}{N}, \quad \text{for} \ \sqrt{ \frac{r\gamma}{N} } \ll 1. \label{eq:dqecslowgam}
\end{eqnarray}
It is thus shown that in the absence of read-out noise ($\gamma = 0$), with only Poisson noise present, a CS recording contains the same amount of Fisher information as a Shannon scan.

In the limit of strong read-out noise (large $\gamma$) $\popt$ in (\ref{eq:sipioptpoisrdta}) approaches $1/2$, and a lower bound is obtained by filling out $p = 1/2$ in the expression for the DQE:
\begin{eqnarray}
  \dqe_{\text{CS}}^{\text{lb}} =  \frac{1}{2N + 4 r \gamma}.
\end{eqnarray}
From this it follows that if $\gamma$ exceeds $(1 - 4 r / N)^{-1} \simeq 1$, a CS reconstruction is preferred since then $\dqe_{\text{CS}}^{\text{lb}} > \dqe_{\text{Sh}}$.  In the limit of strong read-out noise the condition $\gamma > 1$ is met by construction.

Reconstructions from simulations following the settings in Sec. \ref{sec:crbdqesipi} were carried out to test the dependence of the MSE on the relative read-out noise $\gamma$ on CS recordings and Shannon scans.  The optimization ran for 50 iterations, with 10 subiterations for step 1.  For CS, $\gamma$ was varied from $10^{-1}$ to $10^{5}$, and following (\ref{eq:sipioptpoisrdta}) $p$ was set to $\popt$ with values varying from $0.0014$ to $0.45$, respectively.  For the Shannon scans $\gamma$ was varied from $10^{-1}$ to $10^{1}$.  

\begin{figure}
  \center
  \includegraphics[width = 0.99\linewidth]{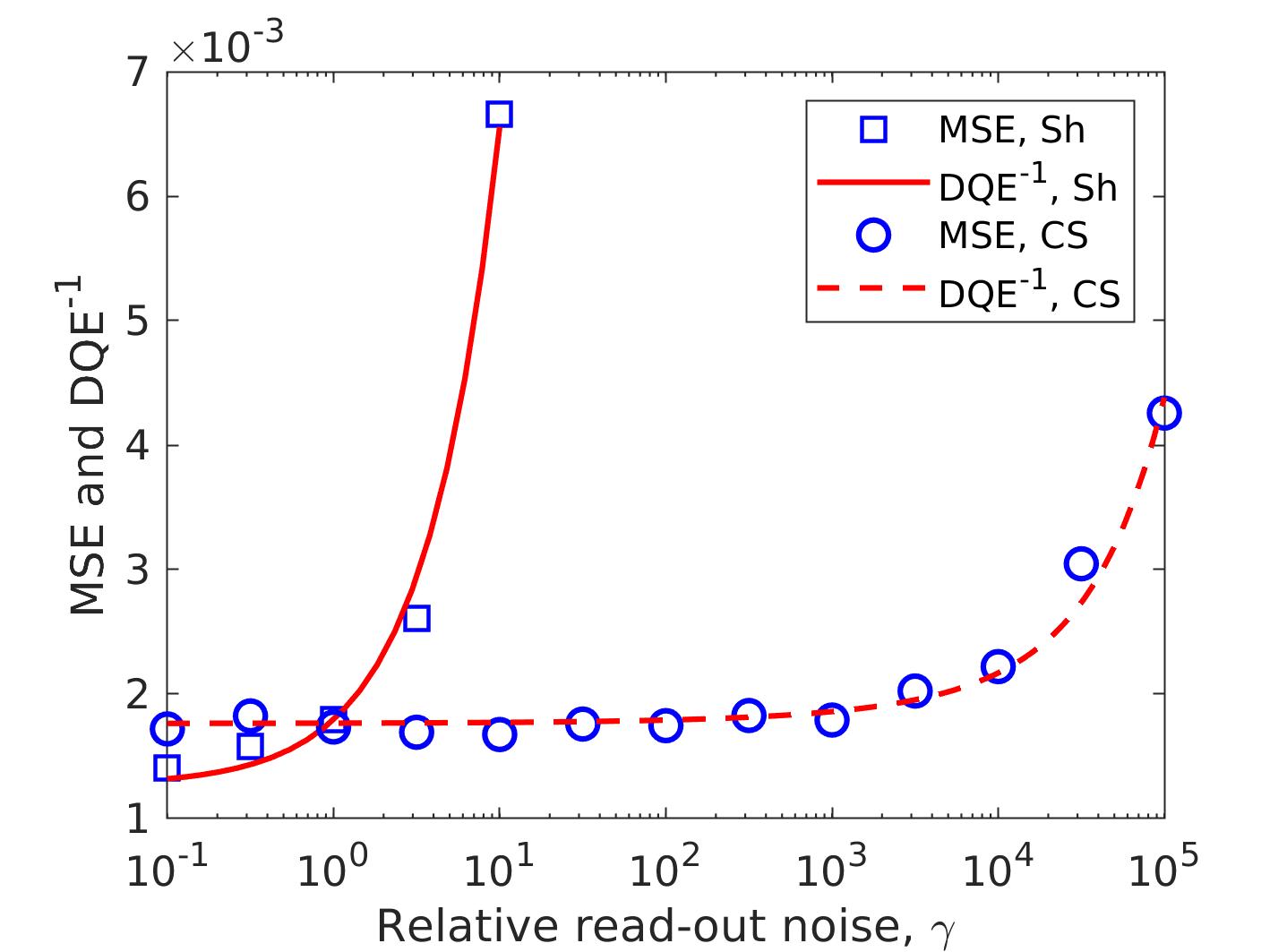}
  \caption{\label{fig:ShCSSipi}  MSE and $\dqe^{-1}$ for reconstructions from Shannon (Sh) scans and from CS measurements for the single-pixel set-up.}
\end{figure}

The results are depicted in Fig. \ref{fig:ShCSSipi}.  After a linear transformation has been applied to $\dqe^{-1}$, it fits well to MSE.  For low noise both approaches yield comparable MSEs, but the much stronger rise for the Shannon scan shows CS is preferable in case of read-out noise.  Also, as predicted in (\ref{eq:dqecslowgam}), MSE for CS is approximately constant for the lower noise levels.

\subsubsection{ADF-STEM}
\label{sec:adfstemre}

The expressions for the $\dqe$s are given in Table \ref{tab:dqes} for ADF-STEM measurements. Contrary to the single-pixel set-up the optimal value for $p$ is independent of the noise model as for all models it holds that,
\begin{eqnarray}
  \frac{\partial}{\partial p} \dqe \propto - \frac{1}{p^2},
\end{eqnarray}
indicating that $\dqe$ is a monotonically decreasing function of $p$. This suggests the optimal value for $p$ is the minimum $1/N$ (see Table \ref{tab:popt}) and that an inpainting approach is preferred. 

The quality of a CS reconstruction is compared to that of a denoised Shannon scan recorded with the same electron dose.  It can be shown that,
\begin{eqnarray}
  \text{if } p \geq \frac{1 + \gamma'/r}{ N } \text{, then } \dqe_{\text{Sh}} \geq \dqe_{\text{CS}}, \label{eq:dqe35}
\end{eqnarray}
where $\dqe_{\text{CS}}$ and $\dqe_{\text{Sh}}$ denote the $\dqe$s for the CS and the Shannon case, defined respectively by sensing matrices \adfcs \ and \adfsh \ for simultaneous Poisson and read-out noise (Table \ref{tab:dqes}, third and fourth row, second column).  

In the absence of read-out noise, $\gamma'$ equals $0$ and condition (\ref{eq:dqe35}) is always met.  $\dqe_{\text{Sh}} \geq \dqe_{\text{CS}}$ then implies that a Shannon scan contains at least as much Fisher information as a CS measurement.  That the respective reconstructions yield approximately the same $\mse$ under equal dose conditions can be inferred from the results in Sec. \ref{sec:expadfstem} for experimental ADF-STEM where read-out noise was negligible.  Since these CS measurements were obtained by selecting $M$ pixels at random from a Shannon scan, the total dose for these CS measurements scales directly with $M$.  The results in Fig. \ref{fig:adfstem2} show that the $\mse$ is a linear function of $\dqe^{-1} = N/M$ and hence of the inverse of the total dose.  This implies that for constant dose $\mse$ is constant as well, and hence equal to that of the denoised Shannon scan at $M/N = 1$.

\section{Discussion}
\label{sec:dis}

Rigorous as the $\crb$ is, it lacks some generality as it yields results for just the test object and the particular realization of the sensing matrix $A$, and requires knowledge of the support in the sparse basis. In contrast, the DQE, although more heuristic in nature, is an analytical function of the experimental conditions and alleviates some of the $\crb$'s drawbacks: the test object enters through just its mean value, only the variance of the sensing matrix entries is needed and knowledge of the support is not necessary.

A caveat is that for too few on-pixels the Fisher information matrix becomes singular, which means no unbiased estimators are possible \cite{li2012} and the $\crb$ cannot be computed.  This is information that the $\dqe$ cannot deliver, and its predictions therefore only hold on the condition of the Fisher information matrix being non-singular. 

That regularized estimators are in general biased is reflected in the fact that a linear fit is needed to match $\dqe^{-1}$ to the MSE of the reconstructions, instead of the mere scaling that sufficed for matching to the $\crb$.  Nevertheless, the $\dqe$ still predicted the optimal experimental settings accurately, thus suggesting that also biased estimators perform best with measurements with maximum Fisher information.

\section{Conclusions}
\label{sec:con}

In this paper the performance of various CS set-ups was assessed under the constraint of constant total illumination and compared to that of a denoised Shannon scan with the same dose.  The A-optimality information criterion $\crb$ was chosen as a measure of the amount of Fisher information in the measurements.  With the aid of simulations, the heuristic quantity detective quantum efficiency $\dqe$ was shown to track the $\crb$ accurately.  Also the mean squared error (MSE) of CS reconstructions from experimental and simulated recordings was well tracked by the $\dqe$.

The $\dqe$'s analytical tractability was then used to show that for the investigated sensing matrices and in the absence of read-out noise, i.e. with only Poisson noise present, compressed sensing does not raise the amount of Fisher information in the recordings above that of a Shannon sampled signal.  As a consequence, reconstruction results yield a comparable MSE when both data sets are treated with the same algorithm. This result holds for both investigated systems, but is particularly surprising for the single-pixel camera, as the Shannon scan makes use of only $1/N$ of the total dose.   This is considered of fundamental importance here as read-out noise can be viewed as a mere engineering problem for particles with sufficiently high energies. This might temper the high expectations for beam damage reduction that were laid out in the TEM community.

The $\dqe$ was also used to optimize the experimental designs, i.e. the fraction $p$ of on-pixels for best reconstruction quality as a function of particle dose, read-out noise and other experimental parameters.  In the presence of Poisson noise these results, summarized in Table \ref{tab:popt}, differ markedly from the optimal $p = 1/2$ often tacitly assumed in literature.  When there is only Poisson noise an inpainting set-up ($p = 1/N$) is optimal in all investigated systems.  No matter the noise model, $p = 1/N$ is optimal for ADF-STEM.  When there is only read-out noise $p = 1/2$ and $p = 2/3$ are optimal if the on-values in the single-pixel camera are discrete or fractional, respectively.  A combination of Poisson noise and read-out noise yields an optimum between these extremes.

\appendix

\section{Expected log-likelihood regularization}
\label{sec:ElnL}

The simulation and experimental results have been obtained by solving the following problem by minimizing the associated augmented Lagrangian \cite{nocedal1999}  through an alternating direction scheme  \cite{boyd2011,li2013,jiang2016},
\begin{eqnarray}
\label{eq:lnLElnLApp}
  \min_x \tv(x) \ \text{ s.t. } \ \ln L( y | x ) = \E \(  \ln L( y | x ) \).
\end{eqnarray}
$\ln L$ is the log-likelihood of the measurements $y$ conditional on the model parameters $x$, and $E$ denotes the expectation value.

\subsection{Expected log-likelihood}

The likelihood $L$ is defined as
\begin{eqnarray}
  L = \prod_{j = 1}^M p( y_j | I A_j x ),
\end{eqnarray}
with $p$ the probability distribution function of the measurements $y$ as given in (\ref{eq:poidis}), (\ref{eq:nordis}) or (\ref{eq:rdtdis}). Filling out expression (\ref{eq:nordis}), yields
\begin{eqnarray}
\label{eq:lnL}
  \ln L = \sum_j -\ln \( I A_j x + c \) - \frac{ \( I A_j x - y_j \)^2 }{ I A_j x + c },
\end{eqnarray}
omitting constant terms and a factor of $1/2$.  Taking the expectation value over $y$ gives
\begin{eqnarray}
\label{eq:ElnL}
  \E \( \ln L \) = \sum_j -\ln \( I A_j x + c \) - \frac{ \sigma_y^2 }{ I A_j x + c }.
\end{eqnarray}
The symbol $\sigma_y^2$ denotes the variance of the measurements, and equals $I A_j x + c$ for Poisson and read-out noise and $c$ for just read-out noise.

Equating $\ln L$ and $\E( \ln L )$ is then equivalent to setting to zero the constraint,
\begin{eqnarray}
  c_{\ln L} = \sum_{j = 1}^M \( \frac{ ( I A_j x - y_j )^2 }{ I A_j x + c } -1 \). \label{eq:clnL}
\end{eqnarray}
The constraint associated with (\ref{eq:rdtdis}) is then
\begin{eqnarray}
	\sum_{j = 1}^M \( \frac{ ( I A_j x - y_j )^2 }{ c } -1 \). \label{eq:cRdt}
\end{eqnarray}

\subsection{Augmented Lagrangian, alternating directions}

The augmented Lagrangian \cite{nocedal1999} then becomes,
\begin{eqnarray}
\begin{split}
  \LA = & \sum_i^{2N} |s_i| - \sum_i^{2N} \lambda_i \( H_i x - s_i \) \\
	      & + \frac{\mu}{2} \sum_i^{2N} \( H_i x - s_i \)^2 - \nu c_{\ln L}\( x \)  + \frac{\beta}{2} c_{\ln L}^2(x). \label{eq:aula}
\end{split}
\end{eqnarray}

This problem is then solved with an alternating directions method \cite{boyd2011,li2013} where in iteration $\ell+1$:
\begin{enumerate}
 \item $\LA$ is minimized w.r.t. $x$ through a numerical optimization with a Polak-Ribi\`ere non-linear conjugate gradient approach \cite{shewchuk1994,nocedal1999}, with $s^{(\ell)}$, $\lambda^{(\ell)}$ and $\nu^{(\ell)}$ kept constant;
\item $\LA$ is minimized analytically w.r.t. $s$ through soft thresholding/shrinkage \cite{lucey2012,li2013},
\begin{eqnarray}
\begin{split}
  s^{(\ell+1)} = & \max \( \absl Hx^{(\ell+1)} - \frac{\lambda^{(\ell)}}{\mu} \absr - \frac{1}{\mu}, 0 \) \\ 
	& \times \text{sgn}\( Hx^{(\ell+1)} - \frac{\lambda^{(\ell)}}{\mu} \),
	\end{split}
\end{eqnarray}
with $x^{(\ell+1)}$ and $\lambda^{(\ell)}$ kept constant;
\item the multipliers $\lambda$ and $\nu$ are then updated through \cite{nocedal1999}
\begin{eqnarray}
  \lambda^{(\ell+1)} & = & \lambda^{(\ell)} - \mu \( H x^{(\ell+1)} - s^{(\ell+1)} \), \\
	\nu^{(\ell+1)} & = & \nu^{(\ell)} - \beta c_{\ln L}\( x^{(\ell+1)} \),
\end{eqnarray}
with $x^{(\ell+1)}$ and $s^{(\ell+1)}$ kept constant.
\end{enumerate}

In the first iteration $\mu$, $\nu$ and all $\lambda_i$ are set to zero and step 1 is iterated until convergence. In this way an initial estimate for $x$ is obtained that obeys the constraint $\ln L = \E(\ln L)$ virtually perfectly. This estimate is then used as the starting point for the subsequent minimization.

During optimization $I A_j x$ can become temporarily negative. To avoid the ensuing numerical difficulties, the respective term in the denominator of (\ref{eq:clnL}) is wrapped in the softplus function
\begin{eqnarray}
  \epsilon \ln \( 1 + \exp\( I A_j x / \epsilon \) \),
\end{eqnarray}
which maps positive variables onto themselves and negative variables to zero, with $\epsilon = c/10$ the width of the transition between both regimes.

\subsection{Convex optimization}

CS reconstruction problems can often be cast into two distinct forms.  The first is the convex optimization problem:
\begin{eqnarray}
  \arg \min_x \( F(x) = | R x - b |_2^2 + \lambda | W x |_1 \), \label{eq:CS1}
\end{eqnarray}
for some positive constant $\lambda$ and sparsifying matrix $W$; $| \ \ |_2$ and $| \ \ |_1$ are the $\ell_2$- and $\ell_1$-norm, respectively.  The second form is:
\begin{eqnarray}
  \arg \min_x \( | W x |_1 \), \text{ s.t. } | Rx - b |_2^2 = C, \label{eq:CS2}
\end{eqnarray}
for some positive target value $C$, chosen for example so as to make the likelihood consistent with the experimental error, such as in (\ref{eq:clnL}).  We assume $C$ is feasible, i.e. there is a non-empty subset $X$ of potential solutions such that $| R x - b |_2^2 = C$ for all $x \in X$.

These two forms are equivalent. As shown in \cite{vandenberg2008siam}, the residual $| R x - b |_2^2$ and the $\ell_1$-norm of the solution, $|Wx|_1$, of (\ref{eq:CS1}) define a convex Pareto-curve when plotted against each other.  And since this curve is parametrized by $\lambda$, $\lambda$ can be used to tune the solution to obtain $| R x - b |_2^2 = C$.  The result so found is also a solution to the second form (\ref{eq:CS2}), for if there were an element of $X$ with lower $ |Wx|_1 $, the first algorithm could have selected it and reduced $F(x)$.

The constraint (\ref{eq:clnL}) is well approximated by a second order expansion in $I A_j x$ around $y_j$ of the individual terms $j$:
\begin{eqnarray}
  c_{\ln L} = \sum_{j = 1}^M \( \frac{ ( I A_j x - y_j )^2 }{ y_j + c } - 1 \). \label{eq:clnLApp}
\end{eqnarray}
This is correct up to second order in $I A_j x - y_j$, and since the summation averages out the third order error terms, the dominant error is only of fourth order.  The translation from the formulation in (\ref{eq:CS2}) to the problem at hand is then 
\begin{eqnarray}
  Rx = \frac{I A x}{ \sqrt{y + c} }, \ b = \frac{y}{\sqrt{y + c}}, \ C = M \text{, and } W = H,
\end{eqnarray}
where division of vectors is elementwise.

\section{Derivation of DQEs}
\label{sec:derdqe}

In this appendix the detective quantum efficiencies, DQE, as defined in (\ref{eq:dqe}) are derived.  For $\snrin$:
\begin{eqnarray}
  \snrin^2 & = & \frac{\( I \sigma \)^2 }{ I \mu } = I \frac{\sigma^2}{\mu} \text{, for single-pixel}, \label{eq:snrinsp} \\
  \snrin^2 & = & I \frac{\sigma^2}{\mu} \frac{M}{N} \text{, for ADF-STEM}. \label{eq:snrinADF}
\end{eqnarray}
where $I \sigma$ is the signal's standard deviation, $I \mu$ is the average number of photons reflecting off each of the object's pixels, and $\mu$ and $\sigma$ are the mean and standard deviation of $x$, respectively.

The SNR of the signal $y$ recorded according to (\ref{eq:linimafor}) follows, 
\begin{eqnarray}
  \snrout^2 = \frac{\sigma_y^2}{\av{y} + c}, \label{eq:snrout}
\end{eqnarray}
where $\sigma_y^2$ is the variance of $y$, $\av{y}$ is---as the average of $y$---the variance of the Poisson noise, and $c$ is the read-out noise variance.

The CS measurement process can be regarded as sampling \emph{without} replacement from the $N$ members of population $x$, and the covariance matrix $\cov_x$ of $x$ hence has diagonal elements $\sigma^2$ and off-diagonal elements $-\sigma^2 / (N-1)$. Since the members of $y$ are measured independently but with the same rules for on- and off-pixels, $\sigma^2_y$ can be taken as the variance of an arbitrary element $k$ in $y$,
\begin{eqnarray}
  \sigma^2_{y} & = & \( \frac{\partial y_k}{\partial x} \)^T \cov_x \frac{\partial y_k}{\partial x}, \\
	               & = &  I^2 \sigma^2 \sum_{i=1}^N a_{ki}^2 - \frac{I^2 \sigma^2}{N-1}\sum_{i=1}^N a_{ki} \sum_{j \neq i}^N  a_{kj}, \\
								 & \simeq & I^2 N \sigma^2 \sigma_a^2.
\end{eqnarray}
with $a_{ki}$ the respective element of sensing matrix $A$ in (\ref{eq:linimafor}), and $\sigma^2$ and $\sigma_a^2$ the variances of $x$ and of the elements in $A$, respectively.

The expressions in Table \ref{tab:dqes} can be obtained by finding $\av{y}$ and $\sigma_a$ for all sensing matrices (see Table \ref{tab:avysig}), and setting to zeros the terms $c$ or $\av{y}$ in the denominator in (\ref{eq:snrout}) for the cases without read-out noise or only read-out noise, respectively. Furthermore, the change of variables in (\ref{eq:varssipi}) helps simplifying the results.

\begin{table}
\renewcommand{\arraystretch}{1.5}
\caption{Average Signals and Variances for Various Sensing Matrices}
\label{tab:avysig}
\center
\begin{footnotesize}
\begin{tabular}{c | c c }
  \hline
	\hline
  sensing\\matrix  & $\av{y}$ & $\sigma_a^2$ \\
  \hline
   {\footnotesize \sipidi }  & $ I \mu p \frac{N}{ M }$   & $ \frac{p}{M^2} \( 1 - p \) $  \\
   {\footnotesize \sipifr }  &  $ I \mu p \frac{N}{ 2M }$ & $ \frac{p}{M^2} \( 1/3 - p/4 \) $   \\
   {\footnotesize \adfcs }  & $I \mu$ & $ \frac{1}{N^2} \( 1/p - 1 \)$  \\
   {\footnotesize \adfsh }  & $ I \mu \frac{M}{N} $ & $ \frac{M^2}{N^3} $  \\
	\hline
	\hline
\end{tabular}
\end{footnotesize}
\end{table}

\section*{Acknowledgment}

W.V.d.B. acknowledges funding from the DFG project BR~5095/2-1 (`Compressed sensing in ptychography and transmission electron microscopy'). The Qu-Ant-EM microscope used for the experimental data was partly funded by the Hercules fund from the Flemish Government. A.B., A.V., and J.V. acknowledge funding from FWO project G093417N (`Compressed sensing enabling low dose imaging in transmission electron microscopy'). C.T.K. acknowledges financial support from the DFG (CRC 951). We thank Prof. Luis Liz-Marzan for kindly providing the Au nanoparticle sample.

\bibliographystyle{unsrt}

\end{document}